\documentstyle[preprint,aps,prb]{revtex}

\newcommand{\mb}[1]{ {\mbox{\boldmath{$#1$}}}  }

\begin{document}
\draft
\title{Coherent Potential Approximation for `d - wave' Superconductivity in 
Disordered Systems.}

\author{A.M. Martin}
\address{D\'{e}partement de Physique Th\'{e}orique,
Universit\'{e} de Gen\`{e}ve, 1211 Gen\`{e}ve 4, Switzerland.}

\author{ G. Litak}
\address{
Department of Mechanics, Technical University of
Lublin, Poland. 
}
\author{ B.L. Gy\"{o}rffy, 
J.F. Annett}

\address{
H.H. Wills Physics Laboratory, University of
Bristol, Royal Fort, Tyndall Avenue, Bristol BS8 1TL, UK.}

\author{ K.I. Wysoki\'nski.}
\address{
Institute of Physics,Maria Curie-Sk\l{}odowska University, Lublin,
Poland.}

\date{\today}
\maketitle

\begin{abstract}
A Coherent Potential Approximation is developed for $s$--wave and $d$--wave 
superconductivity in disordered systems. We show that the CPA formalism 
reproduces the standard pair-breaking formula, the self-consistent Born 
Approximation and 
the self-consistent $T$-matrix approximation in the appropriate limits. 
We implement the theory and compute  
$T_c$ for $s$--wave and $d$--wave pairing using  an attractive nearest 
neighbor  
Hubbard model featuring  both binary alloy disorder and  a uniform 
distribution of scattering site potentials. We determine the density of 
states and examine its consequences for low temperature heat capacity.
We find that our results are in qualitative agreement with measurements
on Zn doped YBCO superconductors.

\end{abstract}


\section{Introduction}

A treatment of disorder is an essential part of the theory of superconductivity.
After all, one must explain why impurity scattering does not cause resistance. 
Thus it is natural that as evidence for novel superconducting states 
multiplies the foundations of the subject, due mainly to Anderson 
\cite{anderson} and Abrikosov and Gorkov\cite{ag,gorkov},  are  being 
re-examined.
The experiments which stimulate most strongly the current revival of interest 
in the problem are those on the high temperature superconductors\cite{bonn}, 
which are now universally regarded as `$d$--wave
superconductors'\cite{annett}, and 
those involving some of the heavy fermion systems which display signs of  
`$p$--wave' pairing\cite{gross}. In what follows we wish to contribute to
the 
theoretical discussion 
\cite{maki,hirschfeld,carbotte,kalugin,varma,lee,balatsky,tsvelik} of the issues 
raised by these very interesting developments.

The case of classic, `$s$--wave', superconductors is by now well
understood. If the 
perturbation does not break time reversal symmetry and the coherence length is 
sufficiently long, so that the pairing potential $\Delta$ does not fluctuate, 
the Anderson Theorem\cite{anderson} guarantees  that there is an absolute gap 
in the quasi-particle spectrum and the main effect of disorder is that the 
density of normal states in the gap equation is replaced by its average over 
configurations\cite{blg1}. On the other hand if the perturbation breaks
time reversal invariance, as is the case with paramagnetic impurities, the 
effect is more dramatic. For instance, the transition temperature $T_{c}$ is 
reduced from its clean limit  value  $T_{c0}$, according to the well known 
pair-breaking formula:

\begin{equation}
\ln{\left( \frac{T_c}{T_{c0}} \right)} =
 \psi{\left(\frac{1}{2} \right)}
 - 
 \psi{\left(\frac{1}{2}+\rho_c \right)} \label{eqkalugin}
\end{equation}     
where $\psi(x)$ is the digamma function and 
$\rho_c = (2 \pi \tau T_c)^{-1}$ is a measure of the strength of the 
scattering and 
$\tau^{-1}$ is the scattering rate\cite{ag,maki}.

By contrast in the case of superconductors whose Cooper pairs are of exotic 
`$p$--wave' or `$d$--wave' character even simple potential scattering,
which does not 
break time reversal symmetry, causes pair breaking\cite{maki}. This fact was 
noted already in the early contributions to the field\cite{larkin64}, but
has 
become a subject 
of intense scrutiny only 
recently\cite{maki,hirschfeld,carbotte,kalugin,varma,lee,balatsky,tsvelik}.
Of particular interest are two dimensional models featuring `$d$--wave'
pairing as 
these may be relevant to experiments on high $T_c$ superconductors.  
Notably,  for cuprates many experiments have explored the variation of $T_c$, 
the density of states and other properties as a function of Ni and Zn 
substitutions on the copper 
sites\cite{weterholt,liang,tsukada,lemberger,cieplak,bernhard,kuo,williams} or 
irradiation 
damage\cite{ginsberg,elesin,kasia}. 
Although a  wide variety of theoretical 
ideas\cite{haas,kampf,haran,hertog,radtke} and 
methods\cite{openov,byers,xu,xiang,abrikosov,blackstead} have  been applied 
to interpret the experiments
a comprehensive picture of the role of disorder 
is far from complete. On a more formal level, an 
intriguing problem arises from the observation of 
  Gorkov and Kalugin\cite{kalugin} that the scattering in models where 
the order parameter has a line of zeros on the Fermi  surface is highly singular 
and this may be a manifestation of interesting new physics. Indeed in 2-d, 
Nersesyan {\it et al.}\cite{tsvelik} predict that the quasi particle density of 
states N(E) approaches zero, even in the disordered state, as power low, 
$\sim|E|^\alpha$, with positive exponent $\alpha$, instead of going to a finite 
value

\begin{equation}
\label{eqwenger}
 N(0) \sim \frac{1}{\gamma}e^{-\frac{1}{\gamma}} 
\end{equation}
where $\gamma$ measures strength of the interaction, as was found by Gorkov and 
Kalugin\cite{kalugin}.
Another interesting and controversial issue is the relative importance of the
 self-consistent Born approximation (SCBA) and resonant scattering
 in the unitarity 
limit\cite{varma,lee}.
Our aim here is to explore the subject systematically on the bases of explicit  
calculations, albeit for a simple, extended Hubbard model with attractive 
interactions and site diagonal randomness only.

In short we will examine the problem of disordered unconventional 
superconductors making use of the coherent potential approximation (CPA). The 
CPA is the most reliable approximation developed for the theory of 
electronic structure of
random metallic alloys in the normal state\cite{hasegawa,elliot}. Notably it 
has been shown to be exact in both the weak and the strong scattering limits, 
and applicable to systems with low as well as high concentration of impurities.
Significantly, the CPA reduces to the self-consistent Born approximation (SCBA)
for weak  scattering impurities, and  agrees with
the self-consistent T-matrix approximation (SCTA) results for strongly 
scattering impurities  of low concentrations. Indeed it remains a good 
approximation in the unitarity limit of resonant scattering\cite{varma,lee}.
Finally, on account of the fact that it becomes exact as the number of nearest 
neighbors goes to infinity, the CPA is often referred to as a mean field theory 
of disorder\cite{volhardt}. 

Given these desirable features it is clearly worthwhile to explore the 
consequences of the  CPA for disordered superconductors. For the case of 
conventional $s$--wave pairing this has already been done, generating many
useful 
results\cite{lustfeld,litak}. Apart from our earlier brief 
report\cite{litak2} and the limited discussion in Ref.
\onlinecite{wenger},
the case of 
superconductors with Cooper pairs of $d$ symmetry will be treated here, within 
CPA,  for the first time.

 We will describe fully our
numerical method, demonstrate that in various limits  our formalism 
reproduces many of the well known results for disordered 
superconductors, and examine in detail the phase diagram
of the local and non-local attractive  two-dimensional  Hubbard models.
In particular we study the variations of $T_c$ with
impurity scattering strength and with impurity concentration
for the case of local $s$--wave pairing as well as non-local 
(extended) $s$--wave and $d$--wave pairing. We also
contrast the cases for a binary alloy, A-B type, disorder 
with the case of uniformly distributed scattering
potentials on each site.  Finally, we investigate the 
DOS, $N(E)$, at  low energies and its consequences for measurements
of the specific heat, comparing our results with
those of Kalugin and Gorkov\cite{kalugin}, Eq. \ref{eqwenger}, and of
Nereseyan, Tsvelik and Wenger \cite{tsvelik}.

\section{Incorporating the CPA into the Bogoliubov de Gennes Equation}

Our starting point is the single band Hubbard model with an attractive extended 
interaction which  is described by the following Hamiltonian 

\begin{equation}
\label{eq:H}
H=\sum_{ij \sigma} t_{ij}c^{\dagger}_{i \sigma} c_{j \sigma} +
\frac{1}{2} \sum_{ij} U_{ij} \hat{n}_i \hat{n}_j - \sum_{i}(\mu-\varepsilon_i)\hat{n}_i
\end{equation}
where $c_{i\sigma}^\dagger$ and $c_{i\sigma}$ are, respectively, the usual
 creation and annihilation operators for electrons on site $i$ with spin 
$\sigma$, and
the local charge operator is 
$\hat{n}_i=\hat{n}_{i\uparrow}+\hat{n}_{i\downarrow}$ with 
$\hat{n}_{i\sigma}=c^\dagger-{i\sigma}c_{i\sigma}$.  
The  chemical potential is $\mu$, $t_{ij}$ are the hopping integrals
(for $i \neq j$) and $\epsilon_i$ is the local site energy. 
The interaction term, $U-{ij}$,  can be either be a  local
attractive interaction ($U_{ii}< 0$) giving rise to $s$--wave pairing,
or a non-local attractive interaction ($U_{ij}<0$ for $i \neq j$)
giving rise to $d$--wave or extended $s$--wave pairing.  Disorder is
introduced into the problem by allowing the site energies, $\epsilon_i$,
to vary randomly from site to site.

Starting from Eq. (\ref{eq:H}) we apply the Hartree-Fock-Gorkov
\cite{Tink75,Dreiz90} 
approximation, which results in the following  Bogoliubov de Gennes 
equation:

\begin{equation}
\label{eq:bdge1}
\sum_l \left( \begin{array}{c} (\imath \omega_n -\varepsilon_i + \mu) 
\delta_{il}+t_{il}~~~~~ \Delta_{ij} \\ 
 \Delta_{ij}^*~~~~~  (\imath \omega_n +\varepsilon_i - \mu) \delta_{il}
-t_{il} \end{array} \right)
 \left( 
\begin{array}{c} G_{11}(l,j;\imath \omega_n)~~  G_{12}(l,j;\imath \omega_n) \\
G_{21}(l,j;\imath \omega_n)~~  G_{22}(l,j;\imath \omega_n) \end{array} 
\right) = 
\delta_{ij} \left(
\begin{array}{c} 1~~  0 \\ 
0~~  1  \end{array} \right),
\end{equation}

\noindent for the Greens function matrix $G(i,j;\imath \omega_n)$ at the 
Mastusbara frequency, 
$\hbar \omega_n=(2n+1)\pi k_B T$. For computational convenience we shall take 
the hopping integral  $t_{ij}$ to 
be non zero only when the sites $i$ and $j$ are nearest neighbors. The 
mean field pairing potential $\Delta_{ij}$ can either be local ($i=j$) or 
nearest neighbor non-local, see Fig. 1, and couples particle and hole 
amplitudes between sites $i$ and $j$. 
Of course, the  above equations are completed by the self-consistency
condition 
that 

\begin{equation}
\label{eq:selfcon1}
\Delta_{ij}= |U_{ij}| \frac{1}{\beta} \sum_n {\rm e}^{\imath \omega_n \eta} 
G_{12}(i,j;\imath \omega_n),
\end{equation}
where $\eta$ is a positive infinitesimal. To 
simplify matters we have assumed that the normal Hartree and exchange terms can 
be absorbed into the definitions of the chemical potential, $\mu$, or the 
hopping integrals $t_{ij}$. As usual, 
Eqs. (\ref{eq:bdge1}) and (\ref{eq:selfcon1}) are to be solved 
subject to the requirement  on the chemical potential that

\begin{equation}
\label{eq:number1}
n=\frac{2}{\beta} \sum_n {\rm e}^{\imath \omega_n \eta} 
 G_{11}(i,i;\imath \omega_n) , 
\end{equation}

\noindent where $n$ is the average number of electrons per unit cell.
Clearly, the Greens function matrix $\mb{G}(i,j;\imath \omega_n)$ determined by 
the above equations depends on  the set of site energies
$\{\varepsilon_i\}$. Our 
task is to find the configurationally averaged Greens function matrix 
$\langle\mb{G}(i,j;\imath \omega_n)\rangle$. Evidently, this is made much easier 
if we assume that the pairing potential does not fluctuate from configuration to 
configuration. As was argued by Gyorffy {\it et al.}\cite{blg1} this is a good 
approximation 
when the T=0 coherence length $\xi_0$ is large. Thus our results will have to be 
treated with appropriate care when applied to superconductors with short 
coherence length such as superconducting cuprates.

Let us now proceed to deploy the CPA strategy for calculating the averaged 
Greens function matrix $\langle\mb{G}(i,j;\imath \omega_n)\rangle$ subject to 
the self consistency conditions:

\begin{equation}
\label{eq:selfcon2}
\overline \Delta_{ij}= |U_{ij}| \frac{1}{\beta} 
\sum_n {\rm e}^{\imath \omega_n \eta} 
\langle G_{12} (i,j;\imath \omega_n) \rangle,
\end{equation}

\begin{equation}
\label{eq:number2}
\overline n= \frac{2}{\beta} \sum_n {\rm e}^{\imath \omega_n \eta} 
\langle G_{11} (i,i;\imath \omega_n) \rangle. 
\end{equation}

The first move in deriving the fundamental equations of the 
 coherent potential approximation is 
to define a coherent medium  Greens function matrix 
$\mb{G^c}(i,j;\imath \omega_n)$ by

\begin{equation}
\label{eq:bdge2}
\sum_l \left( \begin{array}{c} 
(\imath \omega_n + \mu-\Sigma_{11}(\imath \omega_n)) 
\delta_{il}+t_{il}~~~~~
\overline \Delta_{il} \\
\overline \Delta_{il}^*~~~~~ (\imath \omega_n - \mu
-\Sigma_{22}(\imath \omega_n)) 
\delta_{il}
-t_{il} \end{array} \right) \mb{G^c}(l,j;\imath \omega_n)
 = \delta_{ij} \left(
\begin{array}{c} 1~~ 0
\\ 0~~ 1  \end{array} \right).
\end{equation}

 As will be clear latter $\mb{G^c}(i,j;\imath \omega_n) = \langle 
\mb{G}(i,j;\imath 
\omega_n)\rangle$ 
and hence $\Sigma_{11}(\imath \omega_n)$ and $\Sigma_{22}(\imath 
\omega_n)$ are the diagonal components of the usual self-energy.
Note that we did not introduce any off diagonal self-energies 
such as $\Sigma_{12}(\imath \omega_n)$  and $\Sigma_{21}(\imath \omega_n)$
because for  the single site  perturbations of our model they are  
zero. The 
next step is to consider the scattering of the 
quasi-particles, propagating according to 
$\mb{G^c}(i,j;\imath \omega_n)$ by the 
defects described by the potentials:
\begin{equation}
\label{eq:defects}
\mb{V^l}(\imath \omega_n) = \left( \begin{array}{c} 
\varepsilon^l~~~ 0 \\
0~~~ -\varepsilon^l \end{array} \right) - \left( \begin{array}{c}
\Sigma_{11}(\imath \omega_n)~~~
0 \\
0~~~ \Sigma_{22}(\imath \omega_n) \end{array} \right),
\end{equation}
where $l$ labels one of the $m$ different site energies we wish to 
consider.

In a straightforward application of the CPA principles,\cite{elliot}
$\mb{\Sigma}(\imath \omega_n)$ and therefore 
$\mb{G^c}(i,j;\imath \omega_n)$ is determined by the condition 
that these defects do not scatter on the average {\it i.e.}
\begin{equation}
\label{eq:zeroscat}
\sum_{l=1}^m c_l \mb{T^l}(\imath \omega_n)=0
\end{equation} 
where 

\begin{equation}
\label{eq:scatmat}
\mb{T^l}(\imath \omega_n)=
\mb{V^l}(\imath \omega_n)
\left[\mb{1}-\mb{G^c}(i,j;\imath \omega_n)
\mb{V^l}\right]^{-1}
\end{equation}
and
\begin{equation}
\sum_{l=1}^mc_l=1.
\end{equation}
From Eqs. (\ref{eq:zeroscat},\ref{eq:scatmat}) it is now possible, in 
conjunction 
with equations (\ref{eq:selfcon2}-\ref{eq:bdge2}), to calculate 
$\mb{\Sigma}(\imath \omega_n)$ and 
$\mb{G^c}(i,j;\imath \omega_n)$. The numerical methodology for calculating 
$\mb{G^c}(i,j;\imath \omega_n)$ and $\mb{\Sigma}(\imath \omega_n)$  closely 
follows that in Ref.\onlinecite{Mar98} and is 
outlined in 
Appendix A. 

A number of recent studies of superconductors with unconventional pairing 
suggest that the consequences of disorder depend sensitively on the models used 
to describe the randomness.\cite{fehrenbacher,pokrovsky}
Thus   we 
are   going to  investigate  two  different models. The first corresponds to 
binary alloy disorder, where $m=2$. Namely we consider  two {\em types} of
sites 
with site energies  $\varepsilon_1$ 
and $\varepsilon_2$  and concentrations of $c$ and $1
-c$  respectively. The second model is described by   a uniform 
distribution of site 
energies. Here we shall have in mind the 
limit  where $m \rightarrow \infty$ with 
$\varepsilon_l \in [-\frac{\delta}{2}, \, \frac{\delta}{2}]$.
Consequently in 
Eq. (\ref{eq:zeroscat}) the sum $\sum_l$ becomes the integral   
$\frac{1}{\delta}\int 
d\varepsilon_l$.

In the bimodal case, where $m=2$, we can simplify Eq. (\ref{eq:zeroscat}) to 
find 
\begin{equation}
\label{eq:sig11_1}
\Sigma_{11}(\imath \omega_n)=(2c-1)\frac{\delta}{2}-
(\frac{\delta}{2}-\Sigma_{11}(\imath \omega_n))G_{11}^c(\imath \omega_n)
(-\frac{\delta}{2}-\Sigma_{11}(\imath \omega_n))
\end{equation}
where $|\varepsilon_1-\varepsilon_2|=\delta$.

In the second, uniform distribution, case we transform the sum in equation 
(\ref{eq:zeroscat}) into an integral so that 
\begin{equation}
\frac{1}{\delta}\int_{-\frac{\delta}{2}}^{\frac{\delta}{2}} 
T^l(\imath \omega_n) 
d\varepsilon_l=0.
\end{equation}
After some straightforward algebra this leads to 
\begin{equation}
\label{eq:sig11_2}
\Sigma_{11}(\imath \omega_n)
=-\frac{1}{G^c_{11}(\imath \omega_n)}+\frac{\delta}{2}
\frac{1}{\tanh(\frac{\delta G_{11}^c(\imath \omega_n)}{2})}.
\end{equation} 

Thus our CPA calculations will consist of solving numerically either  
Eq. (\ref{eq:sig11_1}) for the bimodal distribution of the site energies, or
Eq. (\ref{eq:sig11_2}) for the case of uniform distribution, to determine the 
self-energies $\Sigma_{11}(\imath \omega_n)$ and $\Sigma_{22}(\imath \omega_n)$.

\section{  Pair Breaking Formula in CPA}

We now relate the CPA formulae derived above to the usual results of 
disordered superconductors, corresponding to the well known pair breaking 
formula Eq. (\ref{eqkalugin}). As is well known\cite{maki} the pair breaking 
formula was 
first derived for  
magnetic impurities in 
$s$--wave superconductors\cite{ag} but it also applies in many other interesting 
circumstances such as our present concern, namely   the case of
non-magnetic 
impurities in 
$d$--wave superconductors\cite{larkin64}.

 To derive it  within the  CPA let us start with the gap equation 
\begin{equation}
\label{eq:gap1}
\Delta_{\vec k}=\frac{1}{N}\sum_{\vec q} U_{\vec k-\vec q} \frac{1}{\beta} 
\sum_{\omega_n} G^c_{12}(\vec q;\imath \omega_n) e^{\imath \omega_n
\delta}.
\end{equation}

As a motivation for our argument we recall the method of Abrikosov and 
Gorkov\cite{ag} for solving the gap equation at $T_c$ for a clean 
superconductor.
  In that case, to find $T_c$ we linearize the analogue of Eq.(\ref{eq:gap1})
 by 
approximating  the off diagonal Greens function  $G_{12}^c$ as follows: 

\begin{equation}
G^{c}_{12}(\vec q;\imath \omega_n) \cong \frac{\Delta_q}
{(\imath \omega_n - \xi_{\vec q})(\imath \omega_n + \xi_{\vec q})}
\label{eq:gapTc}
\end{equation}
where $\xi_{\vec q}=\varepsilon_{\vec q}+\mu$, and for our   tight binding
model with a 
square  lattice  
$\varepsilon_{\vec q}=-2t(\cos(q_x)+\cos(q_y))$. 
Then, we note that the kernel of the linear integral equation for $\Delta(\vec 
k)$ is a four term degenerate kernel:

\begin{equation}
U(\vec k - \vec q) = |U|(\eta_{\vec k}\eta_{\vec q}+\gamma_{\vec k}\gamma_{\vec 
q} + 2 \sin k_x \sin q_x +2 \sin k_y \sin q_y),
\label{eq:pot}
\end{equation}
where 
$\eta_{\vec k}=2(\cos(k_x)-\cos(k_y))$ and $\gamma_{\vec 
k}=2(\cos(k_x)+\cos(k_y))$.  
Consequently, the general $\Delta(\vec k)$ will be a linear superposition of 
$\eta_{\vec k}, \gamma_{\vec k}, \sin k_x$ and $\sin k_y$. However, when the 
internal symmetry of the singlet Cooper pair is pure $d$--wave we may take  
$\Delta(\vec k)$ to be of the form
 \begin{equation}
\Delta_{\vec k}=\Delta_{\eta} \eta_{\vec k}.
\end{equation}
Then the condition for non-zero order parameter  becomes 
\begin{equation}
1=\frac{|U|}{N}\sum_{\vec q} \frac{\eta_{\vec q}^2}{4} T_{c0}
\sum_{\omega_n}\frac{1}
{\omega_n^2-\xi_{\vec q}^2}.
\end{equation}
 Let us now define 
a $d$--wave weighted density of states
\begin{equation}
N_d(E)=\frac{1}{N}\sum_{\vec q} \frac{\eta_{\vec q}^2}{4} \delta(E -
\xi_{\vec q})
\end{equation}
and write the above condition, which determines the transition temperature 
$T_{c0}$, as 
\begin{equation}
\label{eq:transtemp}
1=|U| \int^{\infty}_{-\infty} d \epsilon N_d(E) T_{c0}
\sum_{\omega_n > 0} \frac{2}
{\omega_n^2 + E^2},
\end{equation}
where $\omega_n=\pi T_{c0}(2n+1)$. In the above equation the integral and
the sum are divergent, so we need to introduce a cut-off, $\omega_n^c$. In 
the usual way we assume the density of states $N_d(E)$ is slowly 
varying up to the cut off energy, so we will make the approximation 
$N_d(E)=N_d(0)$. Then, considering that
\begin{equation}
N_d(0)\int^{\infty}_{-\infty} d E 
\frac{1}{\omega_n^2 + E^2}=
\pi N_d(0) \frac{1}{\omega_n}
\end{equation}
we can write 
\begin{equation}
1= |U| N_d(0) 2 \pi T_{c0} \sum_{\omega_n > 0}^{\omega_n^c}
\frac{1}{\omega_n}.
\end{equation}
and hence rewrite Eq. (\ref{eq:transtemp}) as
\begin{equation} 
\label{eq:clean}
\frac{1}{|U| N_d(0)}= 
\psi 
\left(
\frac{1}{2}+\frac{\omega_n^c}{2 \pi T_{c0}} 
\right)- 
\psi
\left(
\frac{1}{2}
\right) \approx
\ln
\left( \gamma \frac{\omega_n^c}{2 \pi T_{c0}} \right).
\end{equation}
This is the BCS result for 
the superconducting transition temperature in the case of $d$--wave 
pairing\cite{Tink75}. It differs from the conventional result only in that the 
$d$--projected density of states $N_d(0)$ has replaced the usual full
density of 
states $N(0)$.

 Let us now return to disordered superconductors and examine how the above well 
known argument is modified when the randomness is dealt with within the CPA.
Using Eq. (\ref{eq:bdge2}) it can be easily seen that instead of 
Eq. (\ref{eq:gapTc})
 we should use
\begin{equation}
G^{c}_{12}(\vec q; \imath \omega_n)=\frac{\Delta_{\vec q}}
{(\imath \omega_n - \xi_{\vec q} - \Sigma_{11}(\imath \omega_n))
(\imath \omega_n + \xi_{\vec q}-\Sigma_{22}(\imath \omega_n))}.
\end{equation}
to linearize Eq. (\ref{eq:gap1}) at $T_c$. Thus, noting that 
\begin{equation}
\Sigma_{22}(\imath \omega_n)=- \Sigma_{11}(-\imath \omega_n)
\end{equation}
the condition which determines $T_c$ can be written as 
\begin{equation}
1=|U| \int^{\infty}_{-\infty} d E N_d(E) T_{c}
\sum_{\omega_n > 0} \frac{2}
{(\imath \omega_n - E - \Sigma_{11}(\imath \omega_n))
(\imath \omega_n + E + \Sigma_{11}(-\imath \omega_n))}.
\end{equation}
Now, at this point we need to know the form of $\Sigma_{11}(\imath \omega_n)$ 
to 
progress any further. As a first approximation we   assume that the most 
important component to the self-energy is the component at the Fermi
energy $E=E_F=\mu$. 
Later on we will test the accuracy of this approximation
by examining our numerical results for $\Sigma_{11}(\imath \omega_n)$. For now, 
however, let us proceed by taking
\begin{equation}
\Sigma_{11}(\imath  \omega_n)= \imath |\Sigma_{0}|sgn (\omega_n). 
\end{equation}
Evidently this leads to 
\begin{equation}
1=|U| \int^{\infty}_{-\infty} d E N_d(E) T_{c}
\sum_{\omega_n > 0} \frac{2}
{(\imath \omega_n - E + \imath |\Sigma_0| sgn(\omega_n))
(\imath \omega_n + E + \imath |\Sigma_0| sgn(\omega_n))}.
\end{equation}
Again taking 
$N_d(E)$  outside of the integration as $N_d(0)$ and performing the integration 
over $E$ we find
\begin{equation}
1=|U| N_d(0) 2 \pi T_{c} \sum_{\omega_n > 0}^{\omega_n^c}
\frac{1}{\omega_n+|\Sigma_0|}
\end{equation}
where 
again  the sum is cut  off, as in the clean limit, by $\omega_n^c$. If we now 
add and
subtract the terms corresponding to $\Sigma_0=0$ (the clean case) we find
\begin{equation}
\label{eq:addsub}
\frac{1}{|U| N_d(0)}=
2 \pi T_{c} \sum_{\omega_n > 0}^{\omega_n^c}
\frac{1}{\omega_n}
+
2 \pi T_{c} \sum_{\omega_n > 0}^{\omega_n^c}
\left(
\frac{1}{\omega_n+|\Sigma_0|}-
\frac{1}{\omega_n}
\right).
\end{equation}
Clearly the term $\frac{1}{|U| N_d(0)}$ on the LHS of Eq. (\ref{eq:addsub})
can be replaced by $\ln \left( \gamma \frac{\omega_n^c}{2 \pi T_{c0}}
\right)$ on account of Eq. (\ref{eq:clean}). With the same accuracy the first
sum on the RHS of equation (\ref{eq:addsub}) equals $\ln \left( \gamma
\frac{\omega_n^c}{2 \pi T_{c}} \right)$ and the second sum is convergent.
Hence the cutoff $\omega_n^c$ can be extended to infinity. As has been noted 
frequently before this infinite sum can be readily performed\cite{ag} and we 
find 
\begin{equation}
\label{eq:pairbreak}
\ln{ 
\left(
\frac{T_c}{T_{c0}}
\right)}
=
\psi
\left(
\frac{1}{2}
\right)-
\psi
\left(
\frac{1}{2}+\rho_c
\right).
\end{equation}
where 
\begin{equation}
\label{eq:rhoc}
\rho_c=\frac{|\Sigma_0|}{2 \pi T_{c}}.
\end{equation}

Eqs. (\ref{eq:pairbreak}) and (\ref{eq:rhoc}) are the central results of this 
section. Reassuringly, whilst equation (\ref{eq:pairbreak}) 
is the standard pair-breaking formula,\cite{maki} Eq. (\ref{eq:rhoc}) is a very 
natural, but novel,  CPA expression for the pair breaking parameter $\rho_c$.
Recall that  our derivation  of the above result from CPA involved  
the approximation: $\Sigma_{11}(\imath \omega_n) \cong \Sigma_0$. To test the 
validity of this 
approximation we wish to compare exact CPA numerical results with the 
predictions of the analytical 
expression: Eqs. (\ref{eq:pairbreak}) and (\ref{eq:rhoc}).
Using numerical solutions of the CPA equation, to be discussed latter,
 Fig. 2 plots the pair breaking strength $\rho_c$ vs. $\delta/t$, 
the disorder strength for the binary alloy type disorder .  
To find pair breaking parameter $\rho_c$  
we calculate $T_c$ for each disorder strength, $\delta/t$, and inverted 
Eq. (\ref{eq:pairbreak}) to obtain an effective $\rho_c$. The exact CPA 
$\rho_c$ can then be compared to the solid line in Fig. 2 where we have 
taken our numerically calculated values for $\Sigma_0$ and directly 
calculated $\rho_c$, via Eq. (\ref{eq:rhoc}). Finally the dashed line in 
Fig. 2 corresponds to  
$\rho_c$ obtained using the Self-Consistent 
Born Approximation (SCBA).  Evidently,  the self energy
at the Fermi energy, $E-\mu=0$, $\Sigma_0$,  gives a good description 
of the pair breaking parameter $\rho_c$ via Eq. (\ref{eq:pairbreak}). 
Also it is clear that, 
as expected, the Self-consistent Born Approximation $\Sigma_0 \equiv 
\frac{\hbar}{\tau}=\pi \delta^2 N(0) $ only works well  in 
the weak scattering limit.

\section{Analytical Features and Predictions  of CPA Equations}

In this section we examine various analytically accessible limits  of the 
CPA formalism described above. Firstly, we demonstrate  that Anderson's theorem 
is 
obeyed for $s$--wave superconductors and the CPA equations are consistent with 
the results of Abrikosov and Gorkov\cite{ag}. Secondly we show that for 
$d$--wave 
superconductors the quasi particle density of states at the Fermi energy, 
$N(0)$, 
is non-zero in the presence of non magnetic disorder scattering 
and is consistent with the
Gorkov-Kalugin formula, Eq. (\ref{eqwenger}).

\subsection{The Anderson's Theorem in the Coherent Potential Approximation}

Formally, the CPA Eqs. (\ref{eq:zeroscat}) ,( \ref{eq:scatmat}) can be written
in 
terms of  renormalized Matsubara frequencies, 
$\tilde\omega_n$, pairing parameter, $\tilde\Delta_{\vec k}$, and particle
energies  
$\tilde\xi_{\vec k}$.
This quantities are defined as follows  
\begin{eqnarray}
\label{eq:renorm}
\tilde \omega_n &=& \omega_n 
\left(1 - \frac{ \imath {\rm Im} \Sigma_{11} (\imath
\omega_n)}{\omega_n}\right) \\
\tilde \Delta_{\vec k}  &=& \bar \Delta_{\vec k} \\
\tilde \xi_{\vec k} &=& \xi_{\vec k} - \mu - {\rm Re} \Sigma_{11} (\imath
\omega_n),
\end{eqnarray}
consequently
\begin{equation}
\label{eq:G_11}
G_{11}(\imath \omega_n)= \frac{1}{N} \sum_{\vec k} 
\frac{\imath \tilde \omega_n + \tilde \xi_{\vec k}}
{ ({\imath \tilde \omega_n})^2 - {\tilde \xi_{\vec k}}^2 - {\tilde
\Delta_{\vec k}}^2}  
\end{equation}
and for the alloy type disorder with $c=0.5$ and $\varepsilon=\pm 
\delta/2$, the 
self energy $\Sigma_{11} (\imath \omega_n)$ which renormalizes 
$\omega_n, \Delta_{\vec k}$ and $\xi_{\vec k}$, and is defined  by
  Eq. (\ref{eq:sig11_1})  can be written as 
\begin{equation}
\label{eq:sig113}
\Sigma_{11}(\imath \omega_n) = 
\frac{ \frac{\delta^2}{4} G_{11}^c (\imath \omega_n)}
{1+G_{11}^c(\imath \omega_n) \Sigma_{11}(\imath \omega_n)}
\end{equation}
 The alternative expression for $\Sigma_{11}(\imath \omega_n)$ in the case of 
   a uniform distribution of local 
potentials, $-\delta/2<\varepsilon^l<\delta/2$ is given in  Eq. 
(\ref{eq:sig11_2}).

Note that in the case of a non isotropic, $d$--wave gap, $\Delta_{\vec k}$, is 
not 
renormalized if the disorder is  diagonal both in site and Nambu space. This is 
different from the case of local $s$--wave pairing where, as can be readily 
shown, $\Delta_{\vec k}$ is renormalized by the same factor as $\tilde\omega_n$ 
in Eq. (\ref{eq:renorm}). Thus for conventional superconductors, in
contradiction to Eqs. (\ref{eq:renorm}) - (\ref{eq:G_11}), we find that the 
CPA yields
\begin{equation}
\label{eq:s_renorm}
\frac{\tilde \omega_n}{\omega_n}=\frac{\tilde \Delta}{\Delta},
\end{equation}
in agreement with Born approximation or Abrikosov-Gorkov theory\cite{ag}. As is 
widely appreciated \cite{ag,blg1} 
the above equation implies the Anderson's Theorem in $s$--wave superconductors. 
By contrast in the $d$--wave case represented by Eqs. (\ref{eq:renorm}), 
Eq.(\ref{eq:s_renorm}) does not hold and hence there is no Anderson theorem.

Finally in concluding this section we would like to stress that Eqs. 
(\ref{eq:renorm}) - (\ref{eq:sig113}) represent strictly pure $d$--wave
result. Even 
if we stick to  the singlet case a more general solution of the CPA equation 
will imply a renormalization of $\Delta_{\vec k}$ to $\tilde \Delta_{\vec k}$.
A good example of such situation is  a case where the symmetry of $\Delta_{\vec 
k}$  is of extended $s$--wave symmetry $s^* \propto (\cos(k_x) +
\cos(k_y))$ type. 
We shall encounter this interesting circumstance later on in this paper.

\subsection{Density of States $N(0)$ in $d$--wave superconductors}

Moving on and returning to the $d$--wave case, we observe that the form of
Eqs.
(\ref{eq:renorm}) - (\ref{eq:sig113}) are the same as were found by 
Larkin.\cite{larkin64} Thus again the CPA reproduces the expected general form 
of the gap and frequency renormalizations, but with an improved description of 
the disorder. The most prominent feature of a conventional superconductor is  
vanishing of the quasi particle density of states $N(\varepsilon)$ for energies 
$\varepsilon$ measured from the Fermi energy $\varepsilon_F$, less then 
$\Delta$. In the case of clean, $d$--wave superconductors the line of
zeros of 
$\Delta(\vec k)$ on the Fermi surface leads to finite $N(\varepsilon)$ for all 
$\varepsilon$ except $\varepsilon = 0$. In fact as is well known\cite{maki} 
$N(\varepsilon)$ approaches zero linearly in $\varepsilon$. In the present 
section we shall investigate what happens to $N(\varepsilon)$ in the presence of 
disorder.

As it turns out for a  given gap parameter $\bar \Delta_k=\Delta \eta_k$ and 
in the limit of 
small disorder $\delta \rightarrow 0$ the CPA equations can be solved
analytically.
To affect the solution  note that in Eq. (\ref{eq:G_11}) the major part 
of the summation is coming from the four singular points in the Brillouin 
zone  
where the denominator vanishes:
\begin{equation}
{\imath \tilde \omega_n}^2 - {\tilde \xi_k}^2 -
{\tilde \Delta_k}^2=0
\end{equation}
Linearizing around these points and performing the summation over $k$
analytically, we find that 
\begin{equation}
G_{11}^c(0)= \frac{ \imath {\rm Im} \Sigma_{11} (0)}
{2 \alpha} \ln \left| \frac{(4 \Delta)^2 + ( {\rm Im} \Sigma_{11}(0))^2}{
({\rm Im} \Sigma_{11}(0))^2 } \right| 
\end{equation}
where $\alpha=2t\Delta \pi$.  
 Clearly in  the limit $|{\rm Im} \Sigma_{11}(0) | << 4 \Delta $
this leads to 
\begin{equation}
\label{eq:G0}
G_{11}^c(0) \approx 
\frac{\imath {\rm Im} \Sigma_{11}(0)}
{\alpha} \ln \left| \frac{4 \Delta}{{\rm Im} \Sigma_{11} (0)} \right|.
\end{equation}
Moreover, when $\Sigma_{11}(\imath \omega_n)$ is small compared to the band 
width we can rewrite Eq. (\ref{eq:sig113}),  using the  
Self-Consistent Born Approximation as 
\begin{equation}
\label{eq:SelfBorn}
\Sigma_{11}(\imath \omega_n)=\frac{\delta^2}{4} G_{11}^c (\imath \omega_n).
\end{equation}
On substituting this result into  Eq. (\ref{eq:G0}) the later becomes an 
equation for $G_{11}^c(0)$ which determines the density of quasi particle state 
$N(0)$ {\it via} the formula
$N(0)=\frac {1}{\pi}G_{11}^c (\imath \omega_n=0)$. Indeed we find 
\begin{equation}
\label{eq:DOS}
N(0) \approx \frac{4\Delta t}{\pi \delta^2} e^{-\frac{8 \pi \Delta t}{\delta^2}}
\end{equation}
and 
\begin{equation}
\label{eq:sig0}
-{\rm Im} \Sigma_{11} (0) \approx 4\Delta t e^{-\frac{8 \pi \Delta
t}{\delta^2}}.
\end{equation}

These formulae agree with the results of Kalugin and Gorkov\cite{kalugin} and  
Haas {\it et al.} \cite{haas} and have been verified numerically. For example 
Figs. 3 and 4  
compare $N(0)$ as calculated from Eq. (\ref{eq:DOS}) and calculated using  
completely self-consistent CPA. As one can see there is good agreement  
between the two results and thus we conclude that at $\varepsilon=0$ the density 
 of states, in CPA,  becomes  finite when an arbitrary small amount of
disorder is 
introduced.  However, one should note that at very low levels of disorder our 
numerical results tend to a constant , whereas the analytical results tend 
to zero, this comes from the small complex component we have added to the energy 
to enable us to evaluate the Greens function. For a more 
detailed description of what is happening to 
$N(0)$ and $- {\rm Im} \Sigma_{11}(0)$, in the presence of disorder, 
see Sec. VII of this paper, where we have analyzed their 
properties more closely and report, extensively,  further  numerical results.

\section{Local Quasi-Particle Density of States Calculations}

In this section we present results for the single site local quasi particle 
density of states 
calculations. As mentioned before we consider two types of disorder, (i) binary 
alloys where we 
have two on-site potentials randomly distributed throughout the lattice and 
(ii) a uniform distribution of random on-site potentials.
For both types of disorder we have  solved, numerically, Eq. 
(\ref{eq:bdge2}) in conjunction with its appropriate self-consistency 
conditions, for the order parameter, $\Delta$, the average number density, $n$,
 and 
the self-energy, $\Sigma_{11}(\imath\omega_n)$.

The first situation we consider is a binary alloy of   random on site 
energies $\varepsilon_1$ and $\varepsilon_2$   with equal concentration 
$c=1-c=0.5$. The parameter we have chosen to use to describe the 
strength of the disorder is $\delta=\varepsilon_1-\varepsilon_2$. Figure
5(a) 
shows the density of states  in the normal state. The 
Van Hove singularity characteristic of a tight-binding model with nearest 
neighbor hopping on a square lattice is  clearly visible for small disorder 
($\delta=0.6t$)  in the middle of the band. As Fig. 5(a) shows, 
for more strongly 
disordered alloys this Van Hove peak is split into two 
peaks with some additional smearing. In the limit of $\delta$ being very 
large we get, as one would expect, band splitting of states associated  with 
$\varepsilon_1$ and $\varepsilon_2$ respectively.
On the other hand the disorder with uniform distribution of site 
energies, 
Fig. 6(a), gives only the smearing and flattening of
density of states with, eventually, a complete   flattening of the
Van Hove peak. 

The imaginary part of the self-energies for these two types of 
disorder are shown in Figs. 5(b) and 6(b) respectively.   
Evidently at low  levels of disorder, they reflect the density of states of 
the clean system, and hence are consistent with the Born Approximation 
{\it via.} 
$-{\rm Im} \Sigma_{11}(E) \propto N(E)$. For larger disorder 
strengths  ${\rm Im}  \Sigma_{11}(E)$
is more flat for the uniform distribution, whereas for the binary alloy the 
self-energy is peaked not at  the Van Hove singularity but at
$\varepsilon_1$ and $\varepsilon_2$.

Before turning to the problem of disordered $d$--wave superconductors for 
reference we have studied, briefly, the $s$--wave case. In short, we have 
introduced a site diagonal, local, attractive interaction with strength $U$, 
into the above model. As expected such interaction leads to conventional
$s$--wave 
pairing. When implementing the CPA 
  we have assumed that in this case the pairing order parameter $\Delta$, which 
is now site diagonal  
does not vary  from site to site. This is consistent with the Anderson 
theorem which was shown to be adequate for systems with a long coherence 
length.\cite{blg1}   
Fig. 7(a) shows the $s$--wave quasi-particle density of states as 
calculated for various 
values of the binary alloy disorder strength $\delta$. 
The results confirm that, for $s$--wave pairing, the 
gap is absolute and, whilst the edges may move, they remain well defined as 
required by    Anderson's Theorem 
\cite{anderson}.  In 
Fig. 7(b) we have also plotted the self-consistent self-energy as a 
function of the quasi-particle energy, for the same disorder strengths 
used to obtain Fig 7(a). From this graph we can see that 
$- {\rm Im} \Sigma_{11}(E) 
$ is zero inside the gap and hence the normal disorder does not act as a pair 
breaker, for $s$--wave superconductivity.

The quasi-particle density of states for  $d$--wave 
superconductors is dramatically different from the above BCS spectrum in the 
$s$--wave 
case even in the clean limit. As it is well known it  has the characteristic 
v-like dip shown in Fig. 8(a), for  $\varepsilon$ near the chemical potential 
($\mu=0$). 
Upon introducing binary  alloy disorder into the 
$d$--wave system one would naturally expect similarly different behavior from 
that described above and in Fig. 7. As is clear from the results reported  in 
Fig. 8 this is indeed the case. Strikingly, as the analytic results of the 
previous section suggested, N(0) becomes non-zero for the slightest disorder. 
This  implies  gapless superconductivity in contrast to the gaped one in the 
$s$--wave case. 

 In Fig. 8(b) 
we have plotted the self-energies, corresponding to the quasi-particle 
density of states results presented in Fig. 8(a). This shows a complex 
evolution with disorder, for small $\delta$ the imaginary part of the
self-energy, $\Sigma(E)$, reflects the 
pure $d$--wave density of states (as expected in the SCBA limit) and hence 
${\rm Im} [\Sigma(0)]\sim 0$. 
However increasing $\delta$ leads to a finite $\Sigma(0)$, 
with a cusp like minimum in ${\rm Im} [\Sigma(E)]$ at $E=0$. 
Increasing 
the disorder even further, to $\delta=2.8t$, the $d$--wave pairing is 
completely destroyed, and ${\rm Im} \Sigma(E)$ 
reverts to the normal system 
self-energy. In this case $-{\rm Im} \Sigma(E)$ 
is a maximum at $E=0$, 
since the Fermi energy, $E_F=\mu$, was set exactly
to $0$.

\section{Critical temperature calculations}

To analyze the effect of the disorder upon $T_c$ we have solved the gap 
equation together with the CPA equations. Again for the sake of comparison we 
have developed the analogous theory for the conventional $s$--wave
superconductors 
based on the site diagonal particle-particle interaction, of strength
$U_0$,
mentioned briefly in a previous sections.
In this case neglecting the 
spatial fluctuations of $\Delta_i$, and linearizing the gap equation for the 
configurationally averaged  single site pairing potential $\overline{\Delta}$ 
we find the condition for the temperature $T_c$, below which   
$\overline{\Delta} \ne 0$, to be
\begin{equation}
\label{eq:tc52}
1= |U_0|
\int_{-\infty}^{\infty} d E 
\frac{\overline{N}_i(E)}{2 E} 
\tanh \left( \frac{\beta_c E}{2}\right)
\end{equation}   
where $\overline{N}_i(E)$ is the averaged density of states in the 
normal state at energy $E$, as calculated within the CPA, and $\beta_c 
=\frac{1}{k_BT_c}$.

The solution of Eq. (\ref{eq:tc52}) for  $T_c$ as a 
function of band filling  and 
strengths of alloyed disorder, $\delta$ is  shown in Fig. 9. From this 
figure we can see that the 
superconducting state exists for all band filling  and the strength of disorder 
does little to suppress $T_c$. Again this is consistent  with Anderson's 
theorem.

Now let us turn to  the case where the interaction is non-local and the pairing 
potential $\Delta_{ij}$ connects nearest neighbors sites. Near $T_c$, 
where the gap equation is linear, the solutions can be labeled by their 
symmetries.  Indeed we find two separate conditions on the temperature
$T_c$ for the 
instability of the normal state to $d$ and extended $s$ symmetry breaking. In the 
first case $T_c^d$ is determined by
\begin{equation}
1=- \frac{|U|}{4\pi} \frac{1}{N}\sum_{\vec k} \int_{-\infty}^{\infty} d E~
{\rm Im} \left(
\frac{\eta_{\vec k}^2 G^c_{11}({\vec k}; E)}
{2 E - \Sigma_{11}(E) -\Sigma_{22}(E)} \right)
\tanh \left(\frac{ \beta_c^d E}{2} \right)
\end{equation}
and in the second extended $s$--wave case, $T_c^s$ is given by
\begin{equation}
1=- \frac{|U|}{4\pi}\frac{1}{N}\sum_{\vec k} \int_{-\infty}^{\infty} d E~
{\rm Im} \left(
\frac{\gamma_{\vec k}^2 G^c_{11}(\vec k;E)}
{2 E - \Sigma_{11}(E) -\Sigma_{22}(E)} \right)
\tanh \left(\frac{ \beta_c^s E}{2} \right),
\end{equation}
where $\eta_{\vec k}$ and $\gamma_{\vec k}$ refer to $s$--wave and
$d$--wave like
`harmonics' defined  
previously following Eq. \ref{eq:pot}.

In Fig. 10 the critical temperature for both $d$--wave and extended $s$--wave
pairing  
is shown as a function of band filling, $n$, for various strengths
of alloyed disorder,  $\delta$. Full lines correspond to $d$--wave $T_c$ while 
dashed ones to extended $s$--wave solutions. In the clean limit 
($\delta=0$) we can see that the extended $s$--wave solution exists mainly 
at the band edges and the $d$--wave solution is largely confined  to the
central portion 
of the band. For the interaction strength we have chosen, the two solutions 
cross at $n \approx 0.38$. Evidently for $n \le 0.38$ the superconducting
instability is at $T_c^s$ whilst $n \ge 0.38$ the transition temperature is
$T_c^d$ As we increase the strength of the alloyed disorder 
($\delta/t=0, 1.2, 2.0, 2.5, 2.6, 2.7, 2.8, 3.0$ from top to bottom curve) we 
see 
that both the $T_c$ for the extended $s$--wave and $d$--wave solutions is 
reduced and for particularly strong disorder ($\delta=2.7t$ and 
$\delta=3.0t$) the maximum in the $d$--wave $T_c$ is no longer at $n=1$.
This is connected with the splitting of Van Hove singularities visible in  
Fig. 5(a), {\it i.e.} the maximum in 
the normal-state density of states corresponds 
to the maximum in the $d$--wave $T_c^d$. 

On the other hand for a uniform 
distribution of random site energies
where there is no splitting of Van Hove singularities, and hence, we see 
a different tendency (Fig. 11, $\delta/t=0, 1.0, 2.0, 3.0, 4.0, 5.0, 6.0$
 from top to bottom curve). The maximum value of $T_c^d$ is 
located 
at $n=1$ no matter what the strength of disorder. For extended $s$--wave 
solutions and for relatively small disorder in the of $d$--wave
case the decreasing tendency in $T_c^{s,d}$
with growing disorder are very similar for the two types of disorder.  
However above a certain strength in either case we note that there is no
crossing between the extended $s$--wave and 
$d$--wave solutions.
This fact may be interpreted as the sign that the $s$ and $d$ sates can
not coexist or mix.

To compare the effect the two types of disorder have upon $T_c$ we have,  
in Fig. 12, calculated $T_c^d$ vs. $\delta/t$ at half filling for both binary 
alloy (solid line) and uniform (dashed line) disorder. From this graph we 
can see that the two types of disorder effect $T_c$ in almost exactly  the 
same manner. 
If we rescale the graphs such that the standard deviation in the disorder 
strength is the same, for uniform disorder $\sigma=\frac{\delta}{2\sqrt{3}}$ 
and for binary disorder $\sigma=\frac{\delta}{2}$, then the two curves lie 
very close to one another.  This prompts the conclusion, which may however 
be premature,\cite{pokrovsky} that $T_c$ is not very sensitive to the
details of the 
fluctuation in the site energy.

Other interesting point to investigate concerns the relative robustness of $T_c$ 
against degradation by disorder in the cases of conventional $s$ and
$d$--wave 
pairing.\cite{abrikosov}
Fig. 13 shows the results for $T_c$ for inter-site $d$--wave and on site 
$s$--wave superconductors 
versus alloyed disorder strength $\delta$, where $n=0.6$. Here one can 
easily recognize a typical difference between these two superconducting 
states.  Clearly   in case of a $d$--wave superconducting state 
disorder acts as an effective pair breaker while for $s$--wave superconductors 
it decreases $T_c$ only slowly or not at all. An interesting physical 
consequence
of this effect is that if both $U_{ii}$ and $U{ij}$ ($i\neq j$) are attractive
increasing disorder  could lead to  a transition from $d$--wave to 
$s$--wave pairing, as suggested by Abrikosov\cite{abrikosov}. 

Evidently, the binary alloy disorder has certain features not shared by the 
uniform distribution of site energies model. One of these is the possibility of 
band splitting.
For a very large $\delta$ parameter the band is split 
into two sub-bands, as shown in Fig. 14.
The on site $s$--wave superconducting state is created within   one of 
sub-bands.
For disorder  parameters: $c=0.5$ and $\epsilon_1=-\epsilon_2=\delta/2$ we 
have not found any similar solution for 
$d$--wave,  but for c=0.25 and n=0.5 a single sub-band $d$--wave solution is 
found. In Fig. 15
we show the very different behavior of $T_c$ for a small and a large 
non-local 
interaction parameter, $U_{ij}$. For large enough $U_{ij}$
the $d$--wave $T_c$ does not go to zero when the band splits. The 
quasi-particle density of states corresponding to such a possibility
is presented in Fig. 16 and, as in Fig. 14, we see the superconducting gap, 
which is now $d$-wave, has formed in one of the sub-bands. 

To complete the 
discussion Figs. 17, 18 and 19 show the critical temperature  plotted 
versus concentration  c
for various strength of binary alloy disorder $\delta$ with band fillings 
of $n=1$ (Fig. 17), $n=0.7$ (Fig. 18) and $n=0.1$ (Fig. 19).
Figs. 17 and 18 correspond to $d$--wave while Fig. 19 correspond to 
inter-site extended $s$--wave pairing. Again for large enough alloy 
disorder pair breaking phenomena take place. If disorder is introduced by a 
fixed potential $\delta$, but varying the concentration of scatters, $c
<< 1$, 
from $0$ we see that the results are similar to the case of a fixed $c$ 
and increasing $\delta$ from $0$. This shows that CPA can work equally 
well, and indeed asymptotically exact, in both the Born (small $\delta$) and 
$T$--matrix (small $c$) and resonant scattering regimes.

\section{ $N(E)$ and  $\Sigma_{11}(E)$  
as  
$E \rightarrow 0$ and the low temperature specific heat}

As mentioned earlier the behavior of the quasiparticle density of states 
$N(E)$ and the imaginary part of the self energy ${\rm Im} 
\Sigma_{11}(E)$ near $E = 0$ is of general conceptual 
significance. For instance, the power law behavior of $N(E) \propto 
|E|^{\alpha}$, for $d$--wave superconductors give rise to power law
dependence 
with temperature of many thermodynamic quantities, such as the specific heat 
$c_v(T)$, instead of the exponential cut--off characteristic of 
a gap in the quasiparticle spectrum.\cite{annett} Naturally, dramatic
changes in the low 
energy behavior of $N(\varepsilon)$ and  ${\rm Im}
\Sigma_{11}(\varepsilon)$ as 
disorder is added to the problem is also of general interest and, as it
turns out, 
of lively
 controversy.\cite{kalugin,varma,lee,balatsky,tsvelik,ziegler,nersesyan}
 In this section we 
wish to examine those predictions of the CPA calculations which are relevant to 
these issues in particular.
  Using the methods outlined in the 
previous sections we calculate  numerically  the quasi-particle  local 
density of states, $N(\varepsilon)$, and the self-energy, 
${\rm Im} \Sigma_{11}(\varepsilon)$, and then  investigate how these two 
quantities change with both temperature and disorder.

  For simplicity we have studied the half filled band ($\mu=0$, 
$n=1$). This is the band filling for which $T_c$, in the $d$--wave case, 
is a maximum.  
We will examine the effects of alloyed disorder on the system and show that 
as we increase the amount of   scattering the specific heat vs. 
temperature relation changes from $T^2$, in the clean limit,
 to $T$ in the limit 
of strong  scattering.  As  is widely recognized \cite{annett}
this behavior is the consequence of $N(E) \sim |E|$ changing to
$N(E)={\rm constant}$ and is consistent with experiments. \cite{hirschfeld}

To get an impression of the form of $N(E)$ and 
$- {\rm Im} \Sigma_{11}(E)$ in the region of the chemical
potential, 
$\mu$, we fit them  to the function $a+b|E|^c$
 in the energy range  
$-\Delta E < E < \Delta E$, where 
$\Delta E$ is small compared to the gap.
Using the coefficients $a$, $b$ and $c$ to fit the curves $N(E)$ 
and $-{\rm Im} \Sigma_{11}(E)$ in the region of $\mu$ gives us 
a tool to analyze their functional form. For example $a$ tells us if the
curves are 
finite at $E=0$, $b$ controls the gradient of the curve and 
$c$ controls the curvature. 

In Figs. 20(a-c) we have plotted these coefficients, for $N(E)$, as a 
function of $T$, for different values of $\delta$. We also included in Fig. 
20(d)  plots of 
the magnitude of the $d$--wave superconducting order parameter $|\Delta|$ 
vs. $T$, for the same values of alloyed disorder strength, $\delta$, 
 to indicate the temperature  where the order parameter goes to zero.
Regarding these curves as a brief summary of what the CPA predicts about the low 
 energy behavior of $N(E)$ and $\Sigma(E)$ we now comment on their 
implications.

Firstly we note that in Fig. 20(a) the parameter $a(T)$
 tends to a finite limit as $T\rightarrow 
0$ and this limit increases more and more rapidly as the disorder described 
by $\delta$, increases.
  Hence   $N(0)$ is finite in agreement 
with our earlier discussion in Sec. IV  
where we derived for low scattering and 
low temperatures the dependence of $N(0)$ on $\delta$ (see Eq. 
(\ref{eq:DOS})). It lends credit to the general 
consistency of our results that the 
$a(T) \cong N(0,T)$ curves rise to their normal state values as $T \rightarrow 
T_c$. Interestingly, as can be seen in  
Fig. 20(b)   the gradient of 
$N(E)$, namely  $b$,  changes dramatically only  near   $T_c$. In fact for the 
case 
where $\delta$ is small (squares) the gradient even  changes sign. This means 
that we go from 
having a gap in the density of states, below $T_c$ to having  
a Van Hove singularity above $T_c$. The curvature, represented by $c$ and 
plotted in Fig. 20(c), shows that for low disorder (squares) as we 
increase the 
temperature we go from a curvature of $c\le 1$, ie. a cusp, to a 
curvature $c > 1$. Finally when the 
critical temperature is reached for each of the four different disorder types 
$c$ goes to $2$.

In the same manner, we now  examine the corresponding coefficients, $a'$,
$b'$ and $c'$ 
for 
$-{\rm Im} \Sigma_{11}(E)$. In Figs. 21(a-c) we have plotted the 
calculated coefficients for $-{\rm Im} \Sigma_{11}(E)$, at different 
values of $\delta$ as a function of the temperature.
In Fig. 21(a) we can see that at low temperatures 
$- {\rm Im} \Sigma_{11}(0)$ increases with disorder, again this 
agrees with our results   in Sec. IV where an analytical form 
for the dependence of  $-{\rm Im} \Sigma_{11}(0)$ upon $\delta$ was 
derived at $T=0$, see Eq. (\ref{eq:sig0}). As we increase the 
temperature we can see 
the self-energy at the chemical potential also rises until the normal state 
value is reached at the critical temperature. In Fig. 21(b) we can see how the 
gradient changes from being large for large $\delta$ and small for small 
$\delta$ at low $T$, and at temperatures greater than $T_c$ to be large 
and negative for large $\delta$ and 
small and negative for small $\delta$. Finally 
Fig. 21(c) shows how, $c$, the curvature goes from being almost linear, 
$c=1$, at low temperatures, regardless of $\delta$,  to $c=2$ above $T_c$.

To illustrate the consequences of these results for the low temperature 
thermodynamic properties we have calculated   the effect of disorder 
upon the specific heat. Using  the above temperature dependent coefficients 
for the density of states near $\varepsilon \sim 0$ we   calculated the limiting 
behavior of the specific heat as $T \rightarrow 0$. The results are shown in 
Fig. 22   for different 
strengths of disorder $\delta$. In the case where $\delta$ is small 
(squares) we can see that the specific heat has a $T^2$ dependence (the 
plotted continuous solid line with no points) and for large $\delta$ the 
dependence upon temperature is linear, as expected.\cite{gross}    
 
\section{Conclusions}
\bigskip

We have compared and contrasted the effect of disorder on conventional
$s$-- and $d$--wave 
superconductors on the bases of an extended, negative $U$ Hubbard model
and a mean field, CPA, treatment of disorder. On the one hand we have derived
 many of the well known results, such as the pair breaking formula in Eq.
(\ref{eq:pairbreak})
or that for the quasi--particle density of states $N(0)$, Eq.
(\ref{eq:DOS}). On the other hand we have 
solved the Gorkov--CPA equations Eqs. (\ref{eq:sig11_1}),
(\ref{eq:gap1}) numerically and
surveyed  the salient features of their consequences by explicit
calculations. The use
 of CPA in describing disordered $d$--wave superconductors 
is an advance in this very active field
\cite{bonn,annett,gross,maki,hirschfeld,carbotte,kalugin,varma,lee,balatsky,tsvelik,fehrenbacher,pokrovsky}
because it allows us to avoid the usually delicate choice between methods,
sets of diagrams,
designed to deal 
with either weakly or strongly scattering perturbations. The CPA treats
both  
kinds of problems equally accurately and it is known to provide a very credible 
interpolation between the two\cite{elliot}. As an example where above feature of CPA
may have a crucial role to play we recall the use of the resonant scatterer
model in interpreting experimental data both on the cuprates
\cite{hirschfeld,kramer}  and
some
Heavy Fermion systems.\cite{varma,Gold98} In short, note that in
the impurity, $c \rightarrow 0$, limit 
the self energy for the Greens function describing an electron moving on a lattice, is given by
\begin{equation}
\Sigma(E) = c T(0,0; E),
\end{equation}
where the $T$--matrix is defined, in terms of the impurity `potential'
$V_{imp}$, a real local shift in the site energy, as 
\begin{equation}
T(0,0; E)= \frac{V_{imp}}{1- V_{imp} G(0,0; E)}.
\end{equation}
Evidently $T(0,0; \epsilon)$ is complex number with an amplitude:

$$
a(E)= \sqrt{(1-V_{imp} {\rm Re} G(0,0;E))^2 + (V_{imp} {\rm
Im}
G(0,0;E))^2}
$$
and a phase, the phase shift $\phi(E)$, given by
\begin{equation}
\tan{\phi(E)} = \frac{V_{imp} {\rm Im} G(0,0; E)}{1-
V_{imp} {\rm Re} G(0,0; E)}.
\end{equation}

Now, observe that while for weak scatterers  
\begin{equation}
\label{eq:sigma_B}
{\rm Im} \Sigma^B(E)= -c \pi \left| V_{imp} \right|^2 N(E) 
\end{equation}
for a resonant scatterer in the unitary limit, defined by $\phi (E)
=\frac{\pi}{2}$, 
\begin{equation}
\label{eq:sigma_R}
{\rm Im} \Sigma^R(E) = -c \frac{\pi}{N(E)},
\end{equation}
\noindent where $N(E)= -\frac{2}{\pi} {\rm Im} G(0,0; E)$. It
is the
above
dramatic difference  in the dependence of ${\rm Im} \Sigma^B$ and ${\rm
Im}
\Sigma^R$
on the density of states $N(E)$ that the cited authors rely on  
in interpreting the relevant experimental data. Evidently, since the individual
 scattering events described by the local $T$--matrices are always treated 
exactly in the CPA,  the CPA describes weak scatterers and resonant scatterers
equally well. Moreover, since it is a reliable approximation not only in the
impurity limit, $c \sim 0$, but also for arbitrary concentrations  it deals
with resonance scatterers even when Eq. (\ref{eq:sigma_R}) no longer
holds.
Thus CPA should be the preferred treatment for models with strong, even resonant
 scattering. 

Although we have not specifically concentrated on this aspect 
of the method, the principle feature of 
Eqs. (\ref{eq:sigma_B}) and (\ref{eq:sigma_R}) , namely the dependence
of $\Sigma (E) $ on $
N( E)$, can be seen to be at work in our calculations of the 
previous section. To demonstrate this we have calculated $\Sigma(0)$ and
N(0) as functions of band filling, $n$, in the most interesting region $n
\sim 1.0$,
and
compared in Fig. 23 their relationships for weak and
strong scatterers at $c=0.5$. Clearly, for weak scatterers
$\delta/t=1$, $-{\rm Im} \Sigma(0)$ is a more or less linear
function
of
$N(0)$ as in Eq. (\ref{eq:sigma_B}), while for strong scatterers
$\delta/t=3$,
${\rm Im} \Sigma(0)$ is inversely proportional to $N(0)$ as in the
resonant
scatterers model described by Eq. (\ref{eq:sigma_R}). Thus we conclude
that
the CPA
employed in the calculation gives a reliable account of disorder in both
the weak and strong scattering regimes. 

Having listed the above desirable
properties of the CPA we hasten to emphasize that it is a `mean--field'
theory of disorder and, hence, does not describe such interesting
phenomena as localization \cite{kramer} even in the normal state. 
\cite{Gold98} Consequently, our result that $N(0)$ is finite for the
smallest amount of disorder can not be taken as evidence against the
conclusion that $N(E) \sim |E|^{\alpha}$ of Nersesyan {\em et al.}.
\cite{tsvelik} As this
originates from the divergence of the vertex corrections in perturbation
series for 
$\Sigma(E)$ we may conjecture that it has to do with localization effects
not described by CPA. This very interesting point is in the need of
further clarification, and indeed we shall return to it in a later publication.

\acknowledgements
This work has been partially supported by EPSRC under grant number GR/L22454 
and KBN grants 2P 03B 031 11 and 2P 03B 050 15. 
A.M. Martin would also like to thank
the TMR network
Dynamics of Nanostructures.

\appendix
\section{The Recursion Method for CPA}
The recursion 
method\cite{RHaydock}  was first used for  solving the Bogoliubov de Gennes 
Equation  by Annett and Goldenfeld\cite{annettgold} and Litak {\it et 
al}\cite{LMG}.
 Its use to implement CPA calculations was advocated  by Julien and 
Mayo\cite{julien}.
 Here, as in reference,\cite{Mar98} we made a combined use of these powerful 
methods to calculate the Greens function matrix
\begin{equation}
\label{eq:6}
G^c_{\alpha \, \alpha^{\prime}}(i,j;E)=
\langle i \alpha|\frac{1}{E \mb{1}- \mb{H}} |j \alpha^{\prime}\rangle
\end{equation}
where the indices $i$ and $j$ denote sites,
while   $\alpha$ and $\alpha^{\prime}$ represent the 
particle or hole
degree of
freedom on each site. We denote particle degrees of freedom by
$\alpha=1$ and hole degrees of freedom by $\alpha=2$. For example
$G^c_{1 \, 2}(i,j; E)$ represents the Greens function between the
particle degree of freedom on site $i$ and the hole degree of freedom
on site $j$ at energy $E$.

To compute the Greens functions (\ref{eq:6}) we can closely follow the method
described by
Martin and Annett.\cite{Mar98}
Using this method we can transform the Hamiltonian to a block
tridiagonal form
\begin{equation}
E \mb{1}- \mb{H}=
\left(\begin{array}{cccccccc}
E \mb{1}-\mb{a_0} & -\mb{b_1} & 0 & 0 & 0 & 0 & 0 & \cdots \\
-\mb{b^{\dagger}_1} & E \mb{1}-\mb{a_1} & -\mb{b_2} & 0 & 0 & 0 & 0 & 
\cdots \\
0 & -\mb{b^{\dagger}_2} & \ddots & \ddots & 0 & 0 & 0 &\cdots \\
0 & 0 & \ddots & \ddots & \ddots & 0 & 0 & \cdots \\
0 & 0 & 0 & -\mb{b^{\dagger}_n} & E \mb{1}-\mb{a_n} & -\mb{b_{n+1}} & 
0 & \cdots \\
\vdots & \vdots & \vdots & 0  & \ddots & \ddots & \ddots & \ddots 
\end{array}
\right)
\label{eq:7}
\end{equation}
where $\mb{a_n}$ and $\mb{b_n}$ are $2 \times 2$ matrices.  Given this
form for $\langle i \alpha|E \mb{1}- \mb{H}|j \alpha^{\prime}\rangle$
and expressing the Greens function as
\begin{equation}
G^c_{\alpha \, \alpha^{\prime}}(i,j; E)=
\langle i \alpha|(E \mb{1}- \mb{H})^{-1} |j \alpha^{\prime}\rangle,
\label{eq:8}
\end{equation}
the Greens functions above can be evaluated as a matrix continued
fraction so that
\begin{equation} 
\mb{G}(i,j;E)= 
\left(E \mb{1}-\mb{a_{0}}-\mb{b^{\dagger}_1} 
\left(E \mb{1}-\mb{a_{1}}-\mb{b^{\dagger}_2} 
\left(E \mb{1}-\mb{a_{2}}-\mb{b^{\dagger}_3} 
\left(E \mb{1}-\mb{a_{3}}-\ldots \right) ^{-1} \mb{b_3} 
\right) ^{-1}\mb{b_2} \right) ^{-1}\mb{b_1} \right) ^{-1} 
\label{eq:9}
\end{equation}
where
\begin{equation}
\mb{G^c}(i,j;E)=\left(
\begin{array}{cc}
G^c_{\alpha \, \alpha}(i,i;E) & 
G^c_{\alpha \, \alpha^{\prime}}(i,j;E) \\
G^c_{\alpha^{\prime} \, \alpha}(j,i;E) & 
G^c_{\alpha^{\prime} \, \alpha^{\prime}}(j,j;E)
\end{array}
\right).
\label{eq:10}
\end{equation}

Within Eqs. (\ref{eq:7}) and (\ref{eq:9}) we have a formally exact
representation of the Greens functions.  However in general 
both the tridiagonal
representation of the Hamiltonian, and the 
matrix continued fraction 
(\ref{eq:7}) will be infinite.  In practice one can only calculate
a finite number of terms in the continued fraction exactly.
In the terminology of the recursion method
it is necessary to {\em terminate} the continued fraction.
\cite{RHaydock,LMG,AMagnus,DTT,CMMNex,GAllan,TKW}

If we were to calculate up to
and including $\mb{a_n}$ and $\mb{b_{n}}$ and then simply set subsequent
coefficients to zero
then the Greens function
would have $2n$ poles along the real axis.  The density of states
would then correspond to a set of $2n$ delta functions. In order to 
obtain accurate results it would be necessary to calculate a large number
of exact levels, which would be expensive in terms of both computer
time and memory.

As a more efficient alternative we choose to terminate the continued fraction 
using the extrapolation method, as used previously by Litak, Miller and 
Gy\"orffy.\cite{LMG} We calculate the values for $\mb{a_n}$ and $\mb{b_n}$ 
exactly up to the first $m$ coefficients using the recursion method. Then, 
noting the fact that the elements of the matrices $\mb{a_n}$ and $\mb{b_n}$ 
vary in a predictable manner,\cite{LMG} we extrapolate the elements of the 
matrices for a further $k$ iterations, where $k$ is usually very much greater 
than $m$. This enables us to compute the various Greens functions accurately 
with relatively little computer time and memory.

We now have a method for calculating the Greens Function 
$G^c_{\alpha \, \alpha^{\prime}}(i,j; E)$. To calculate our system of 
equations self-consistently for a given site 
$i$, we need to know $G^c_{1 \, 2}(i,j_1;E)$, 
$G^c_{1 \, 2}(i,j_2;E)$, $G^c_{1 \, 2}(i,j_3;E)$ and 
$G^c_{1 \, 2}(i,j_4;E)$. If the interaction 
is 
non-local, else if the interaction is local then we only need to calculate 
$G_{1 \, 2}(i,i;E)$, see Fig. 1.  To 
calculate $\overline{n}$ we also need $G^c_{1 \, 1}(i,i; E)$.

In performing our self-consistency calculation we can make use of the fact 
that $\mb{\Sigma}(E)$ just acts as a renormalization to $E 
\mapsto \tilde E$ 
(see Sec. IV). This means we do not need to recalculate $\mb{a_n}$ and 
$\mb{b_n}$ when we calculate the self-consistent self-energy. We first 
calculate the order 
parameter self-consistently for a given attractive interaction, then we 
calculate $\mb{G^c}(i,i; E)$ and hence the self-energy 
self-consistently (remembering that in this self-consistent loop we do not 
need to recalculate $\mb{a_n}$ and $\mb{b_n}$, since the self-energy does 
not affect them). Then we recalculate the superconducting order parameter 
self-consistently, incorporating the calculated self energy and repeat this 
procedure until both the order parameter and the self energy have reached 
stable self-consistent solutions.

\begin{figure}
\caption{Schematic diagram of a tight binding lattice with particle 
and hole degrees of freedom. The gap parameter
$\Delta_{ij}$ couples particles on site $i$ to 
holes on site $j$. The difference between local and non-local pairing is 
highlighted by the dashed (local pairing) and solid (non-local pairing) lines.}
\end{figure}

\begin{figure}
\caption{The effective pair-breaker $\rho_c$ as calculated (i) by numerically 
finding $\frac{T_c}{T_{c0}}$  and inverting Eq. (36) (squares), (ii) 
numerically finding ${\rm Im} \Sigma_0$ and using this in Eq. (37) (solid
line) 
and (iii) Using the self-consistent Born approximation to find
${\rm Im} \Sigma_0$ and 
then 
evaluating Eq. (37) (dashed line).}
\end{figure}

\begin{figure}
\caption{A comparison of the density of states at the chemical potential vs. 
different strengths of alloyed disorder $\delta/t$. The solid line is the 
analytical form derived in Eq. (48) and the dashed one  represent are our 
self-consistent numerical calculations.}
\end{figure}

\begin{figure}
\caption{A comparison of the density of states at the Fermi energy vs. 
different strengths of disorder with a uniform distribution between 
$-\frac{\delta}{2}$ and $\frac{\delta}{2}$. The solid line is the analytical 
form derived in Eq. (48) and the dashed one represent our numerical 
calculations.}
\end{figure}

\begin{figure}
\caption{(a) Density of states $N(E)$ and, (b), self-energies 
${\rm Im} \Sigma(E)$
for a normal state with various alloyed disorder strengths $\delta/t$, 
calculated for a local interaction $|U|=3.5t$ and $n=1$.}
\end{figure}

\begin{figure}
\caption{(a) Density of states $N(E)$ and (b) self-energies 
${\rm Im} \Sigma(E)$ for 
a normal state with various uniform disorder strengths  $\delta/t$, 
calculated for a local interaction $|U|=3.5t$ and $n=1$.}
\end{figure}

\begin{figure}
\caption{(a) Density of states $N(E)$ and (b) self-energies 
${\rm Im} \Sigma(E)$ for a
superconducting state with various alloyed disorder strengths $\delta/t$, 
calculated for a local interaction $|U|=3.5t$ and $n=1$.}
\end{figure}

\begin{figure}
\caption{(a) Density of states $N(E)$ and (b) self-energies 
${\rm Im} \Sigma(E)$ for a
superconducting state with various uniform disorder strengths $\delta/t$, 
calculated for a non-local interaction $|U|=3.5t$ and $n=1$.}
\end{figure}

\begin{figure}
\caption{Critical temperature T$_c$ vs. band filling $n$ for a local  
$s$--wave superconductor with various alloyed disorder strengths $\delta/t$, 
calculated for a local interaction $|U|=3.5t$).}
\end{figure}

\begin{figure}
\caption{Critical temperature for $d$--wave and extended $s$--wave order 
parameters as function of band filling for different strengths of alloyed 
disorder $\delta/t$, calculated for a non-local interaction $|U|=3.5t$.}
\end{figure}

\begin{figure}
\caption{Critical temperature for $d$--wave and extended $s$--wave order 
parameters as a function of band filling for different strengths of uniform 
disorder $\delta/t$, calculated for a non-local interaction $|U|=3.5t$.}
\end{figure}

\begin{figure}
\caption{Critical temperature of the $d$--wave superconducting state vs. 
the strength of disorder. The solid line is for a uniform distribution of 
disorder and the dashed line is for a binary alloy, where $c=0.5$ and in both 
cases $n=1$ and the non-local interaction  $|U|=3.5t$.}
\end{figure}

\begin{figure}
\caption{Critical temperature of $d$--wave (solid line) and local 
$s$--wave (dashed line) superconducting pairing states as a function of 
alloyed disorder for $|U|=3.5t$, $n=0.6$ and $c=0.5$.}
\end{figure}

\begin{figure}
\caption{Local  density of states, at zero temperature, for an alloyed 
disorder of $\delta/t=6$, a local interaction $|U|=3.5t$, $n=0.6$ and $c=0.5$.}
\end{figure}

\begin{figure}
\caption{Critical temperature of the $d$--wave superconducting state as a 
function of alloyed disorder strength shown for different interaction strengths, 
$c=0.25$ and $n=0.5$.}
\end{figure}

\begin{figure}
\caption{Local particle density of states, for an alloyed disorder of 
$\delta/t=8$, a non-local interaction $|U|=3.5t$, $n=0.5$ and $c=0.25$.}  
\end{figure}

\begin{figure}
\caption{Critical temperature vs. concentration, at different alloyed 
disorder strengths, for a non-local interaction $|U|=3.5t$ and $n=1.0$.}
\end{figure}

\begin{figure}
\caption{Critical temperature vs. concentration, at different alloyed disorder 
strengths, for a non-local interaction $|U|=3.5t$ and $n=0.7$.}
\end{figure}

\begin{figure}
\caption{Critical temperature vs. concentration, at different alloyed 
disorder strengths, for a non-local interaction $|U|=3.5t$ and $n=0.1$.}
\end{figure}

\begin{figure}
\caption{The coefficients $a$ (a), $b$ (b) and $c$ (c) which represent the 
Local Particle Density of States in the region of the chemical potential 
as a function of temperature, for different alloyed disorder strengths 
$\delta$ ($\delta=0.6t$ (diamonds), $\delta=1.0t$ (+), $\delta=2.0$ (squares) 
and 
$\delta=2.6t$ ($\times$)). In (d) we have also plotted the magnitude of the 
$d$--wave gap vs. temperature, for the same  alloyed disorder strengths as in 
figures (a), (b) and (c). In all four figures the interaction is non-local
with  
$|U|=3.5t$, $n=1$ and $c=0.5$.}
\end{figure}   

\begin{figure}
\caption{The coefficients $a'$ (a), $b'$ (b) and $c'$ (c) which represent  
the imaginary part of the self-energy, in the region of the  Fermi energy, as 
a function of temperature, for different alloyed disorder strengths $\delta$ 
($\delta=0.6t$ (diamonds), $\delta=1.0t$ (+), $\delta= 2.0t$ (squares) and 
$\delta=2.6t$ ($\times$)). The interaction is non-local with 
$|U|=3.5t$, $n=1$ and $c=0.5$.}
\end{figure}

\begin{figure}
\caption{The specific heat as a function of temperature for different 
strengths of alloyed disorder $\delta$ ($\delta=0.6t$ (squares), $\delta=1.0t$ 
(circles), $\delta=2.0t$ (triangles) and $\delta=2.6t$ (+)), again the
interaction is non-local with $|U|=3.5t$, $n=1$ and $c=0.5$.}
\end{figure}

\begin{figure}
\caption{Relations between ${\rm Im} \Sigma(0)$ and $N(0)$ for weak
$\delta/t=1.0$ an strong $\delta/t=3.0$ scatterers in  binary alloys
 ($c=0.5$).}  
\end{figure}

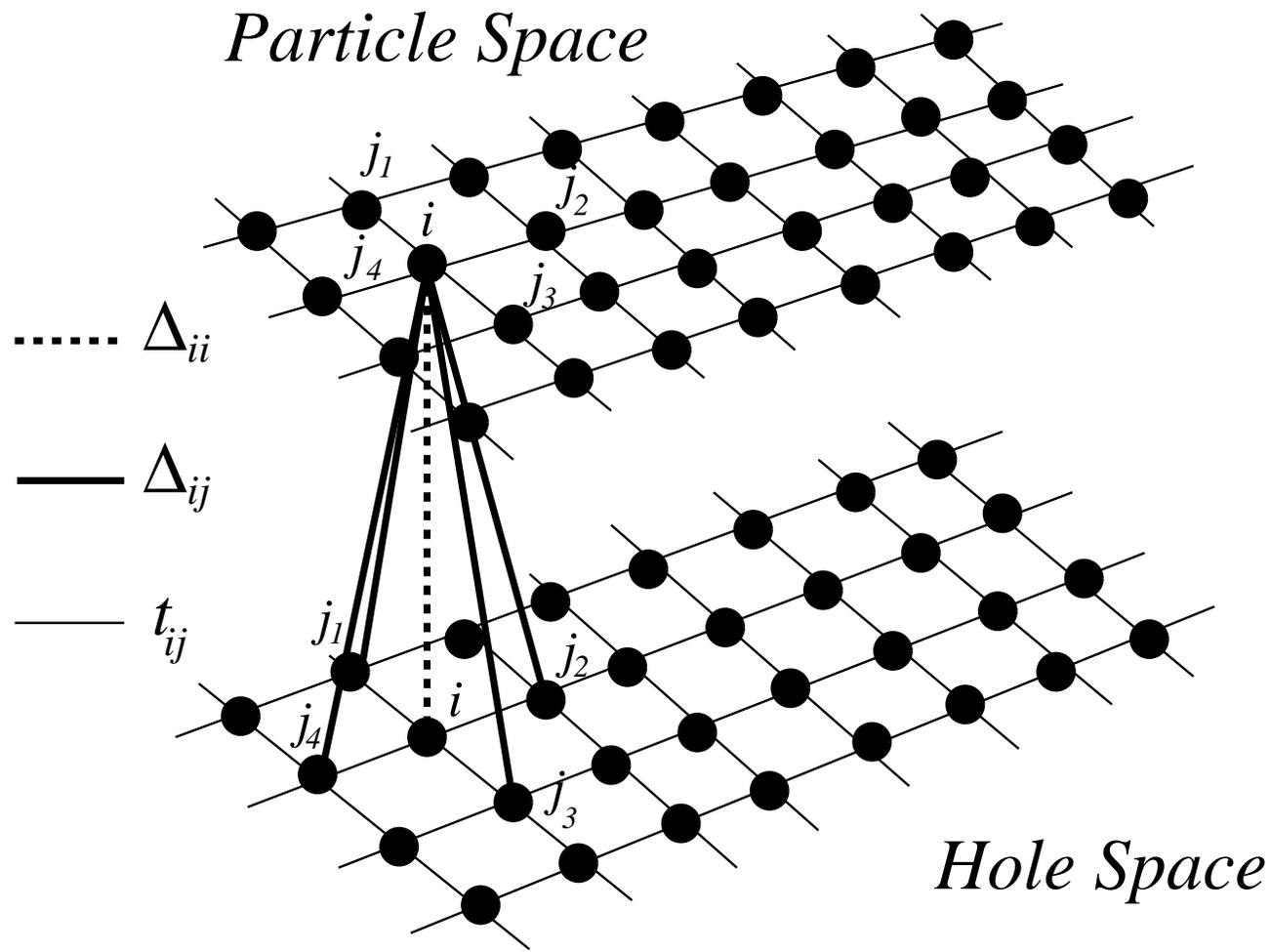
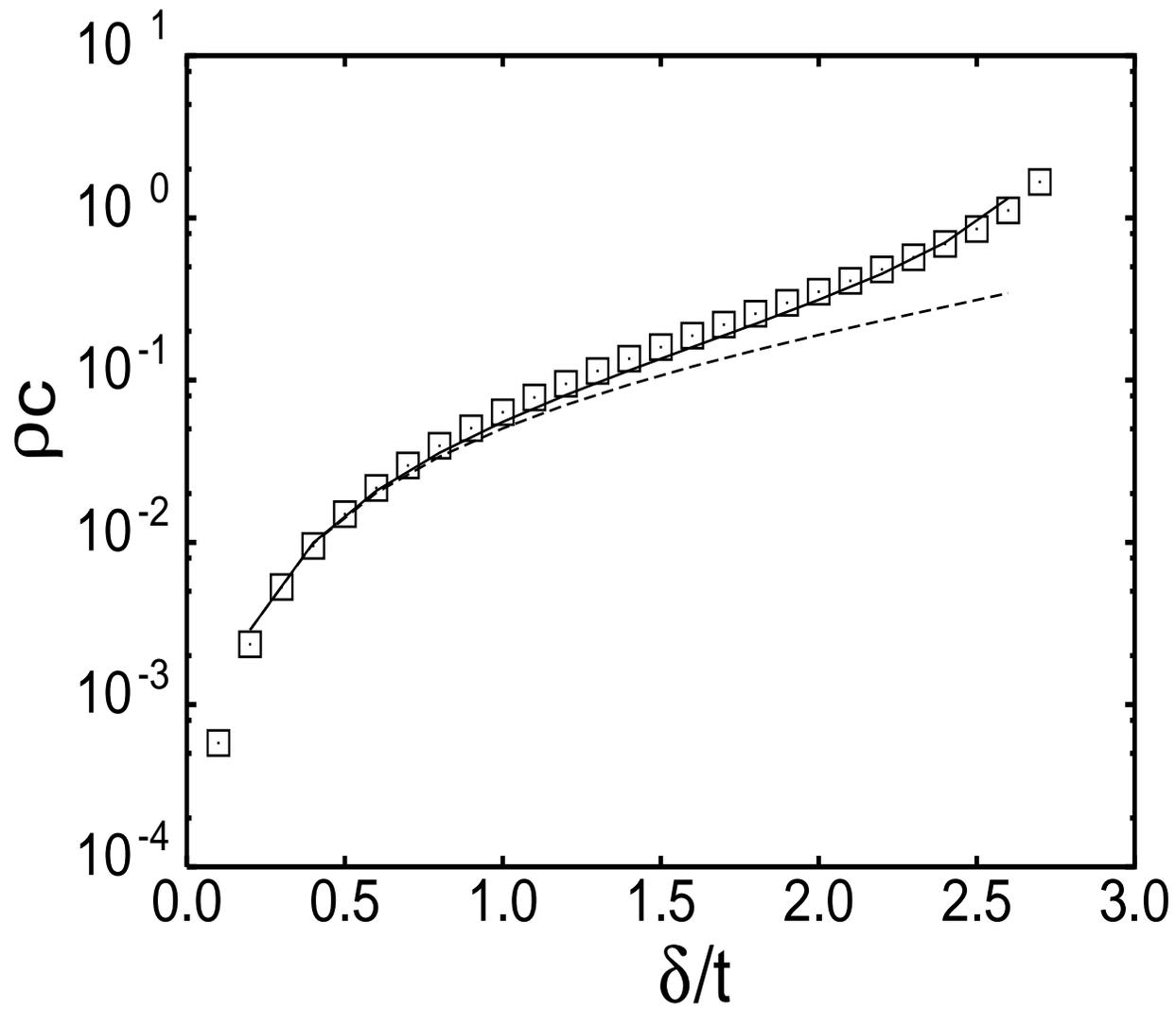
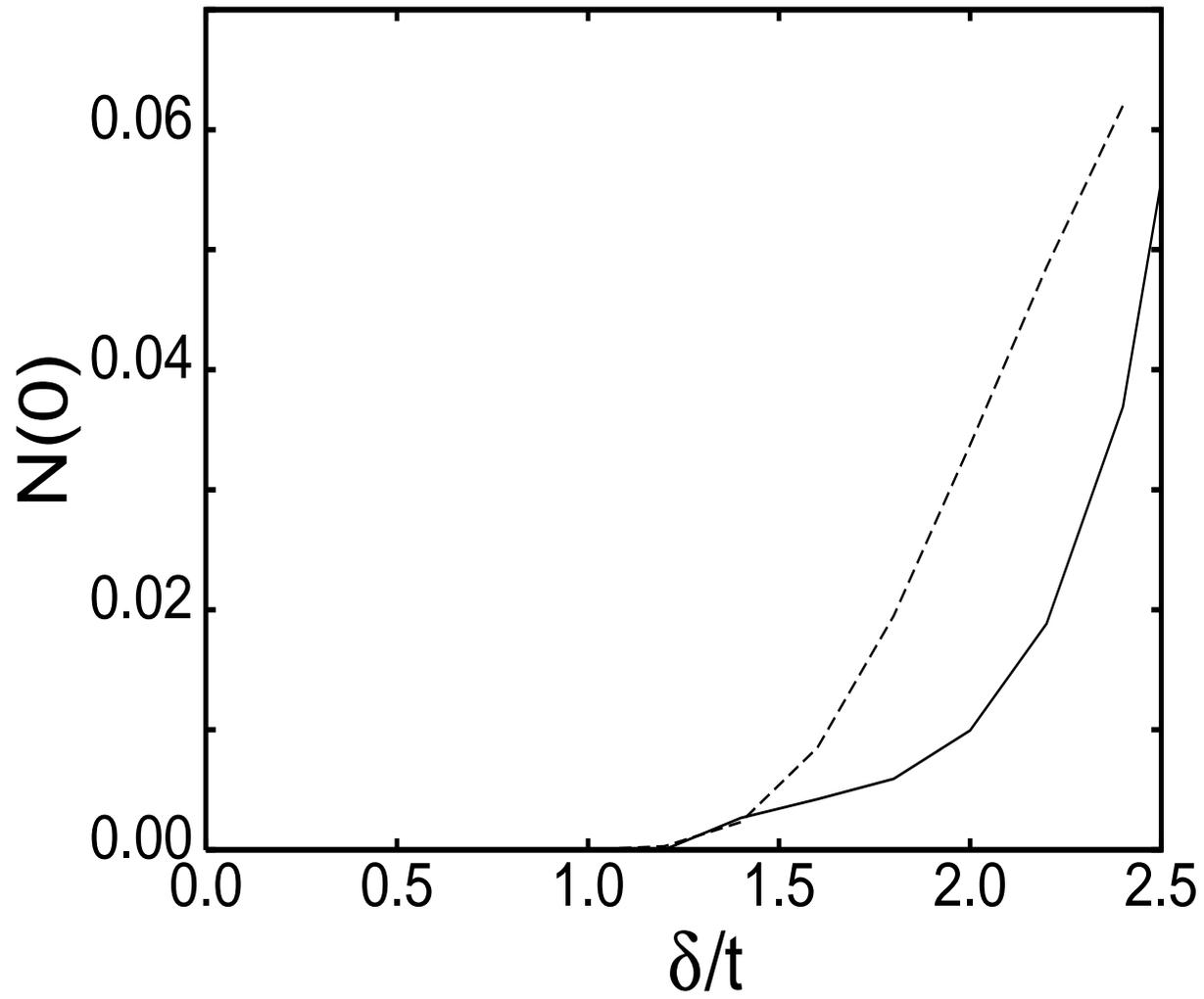
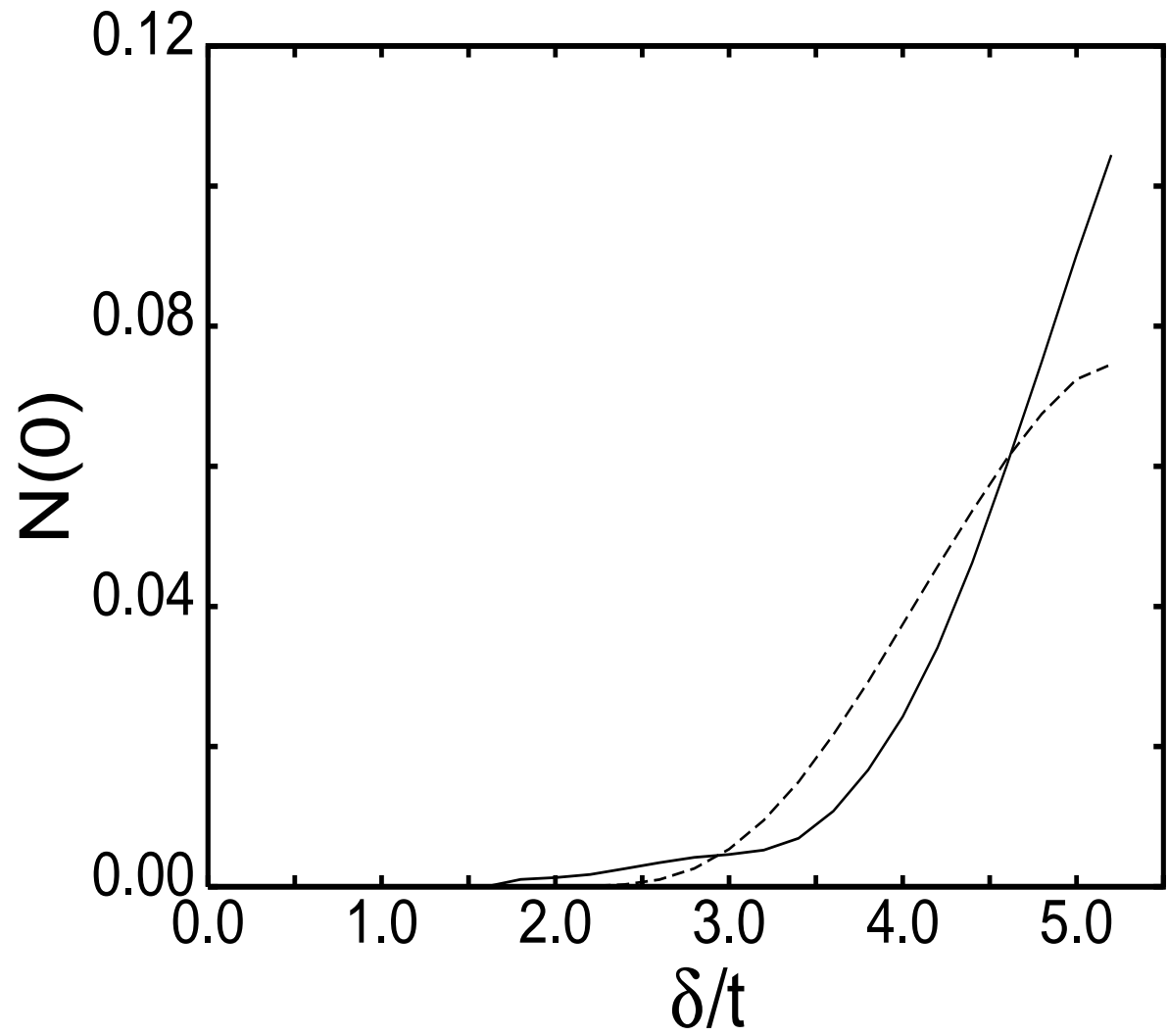
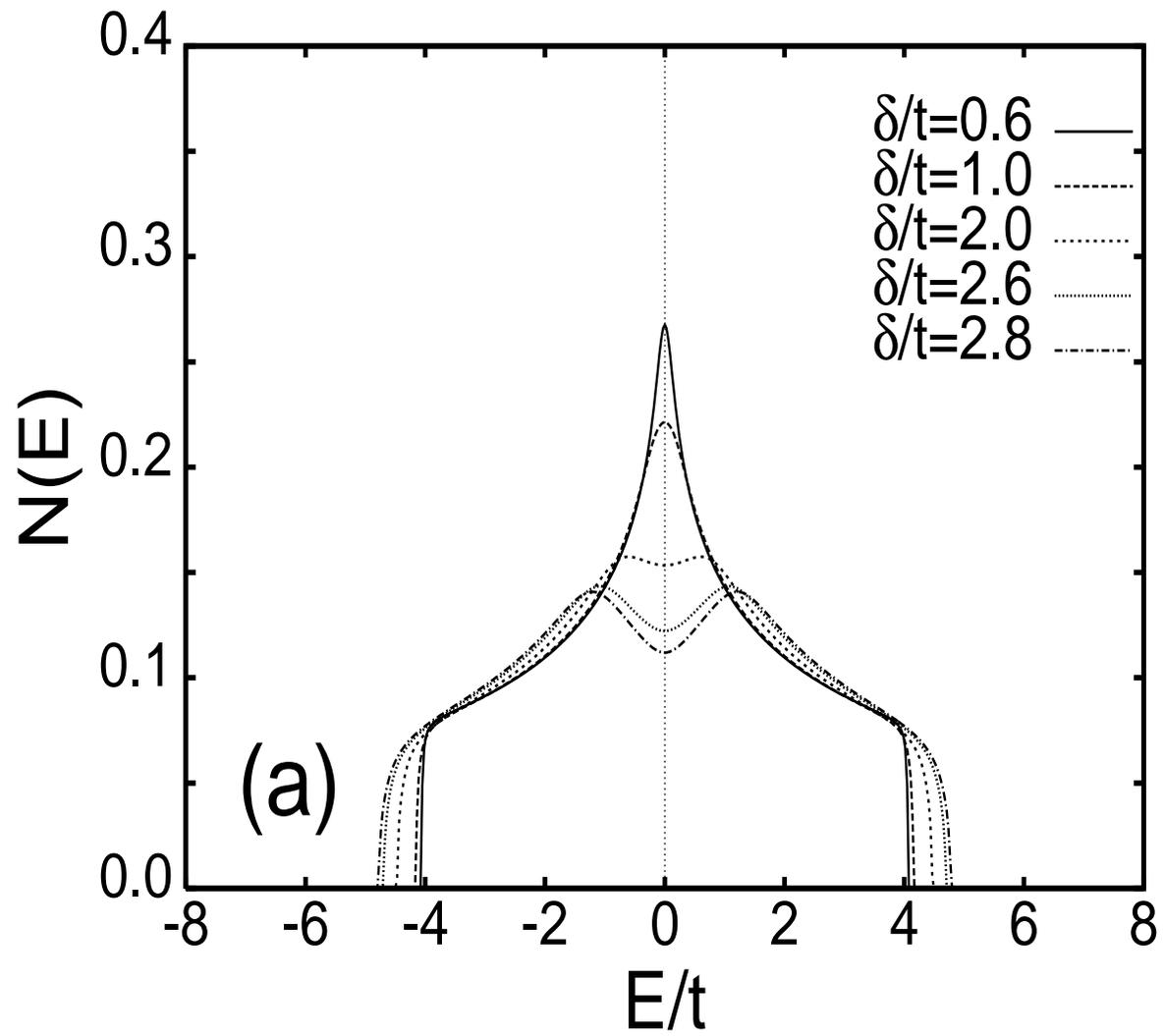
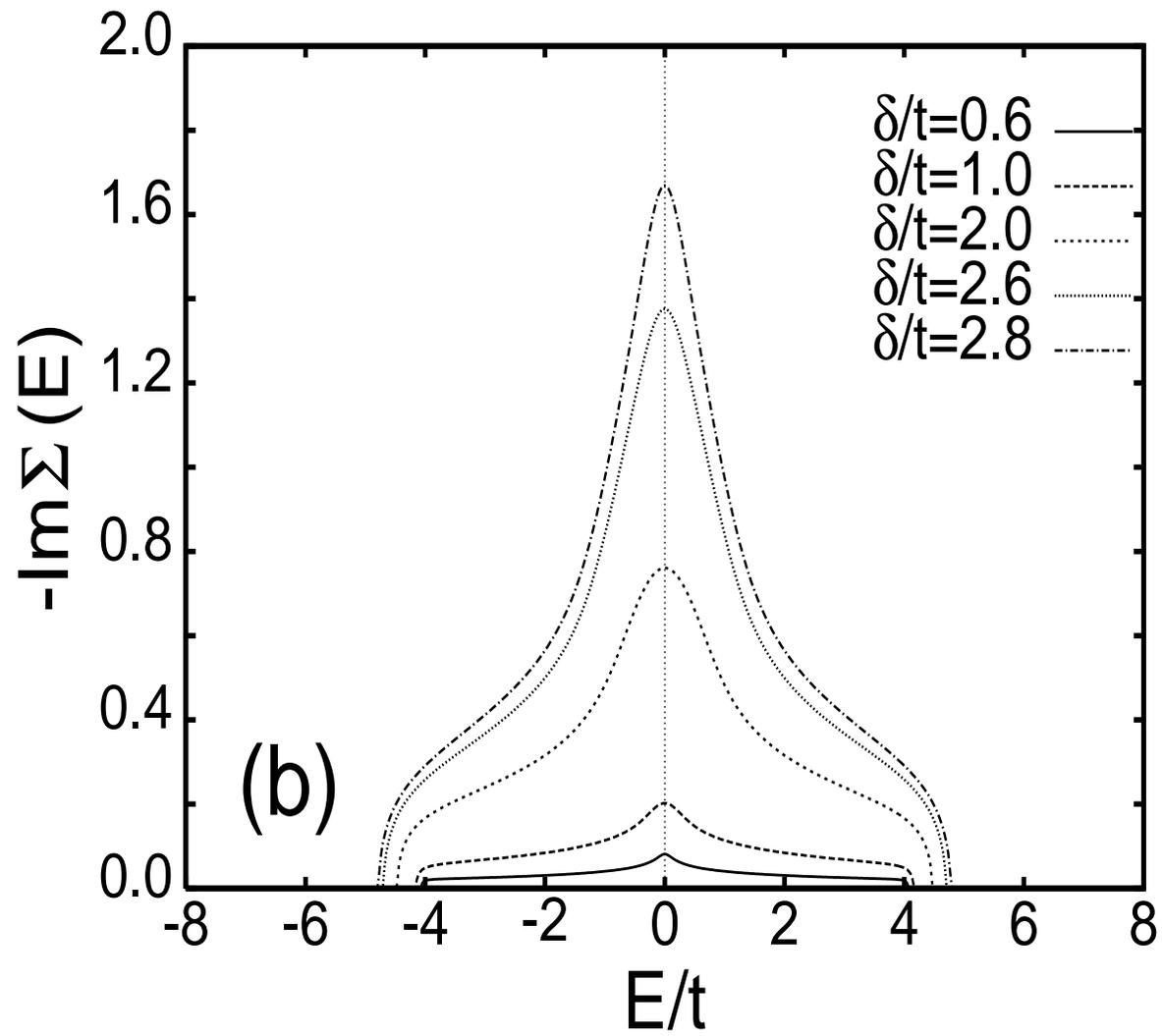
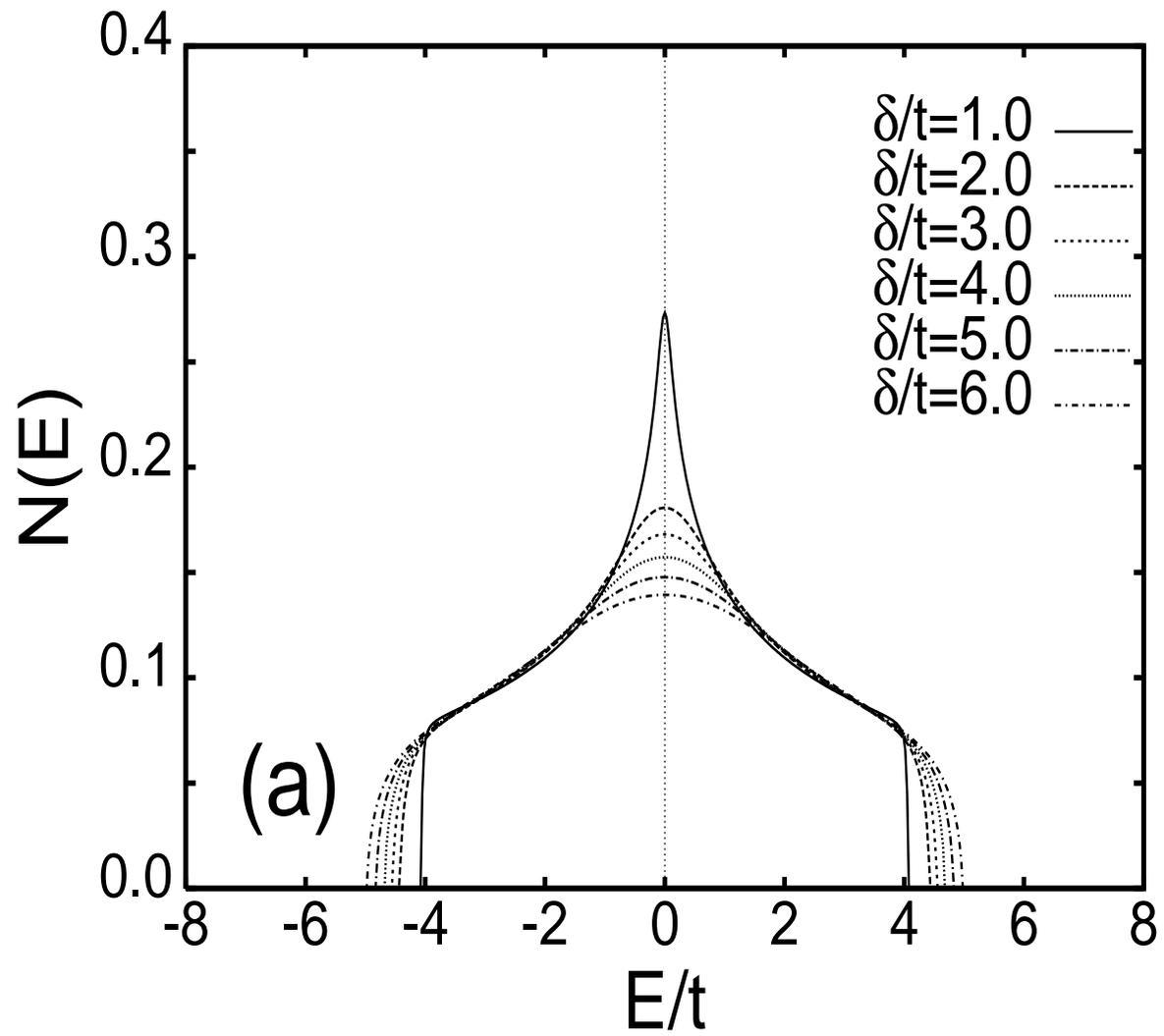
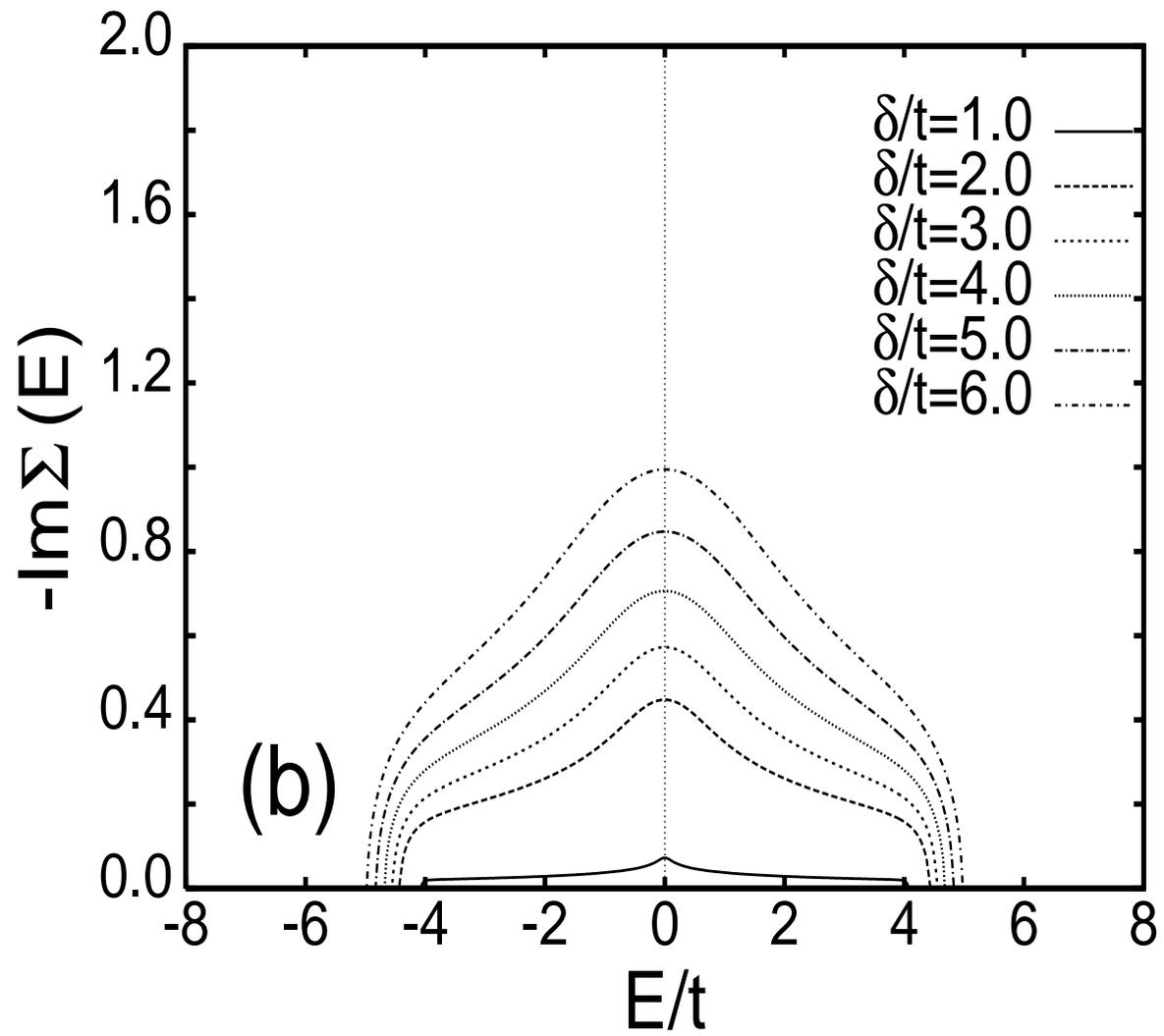
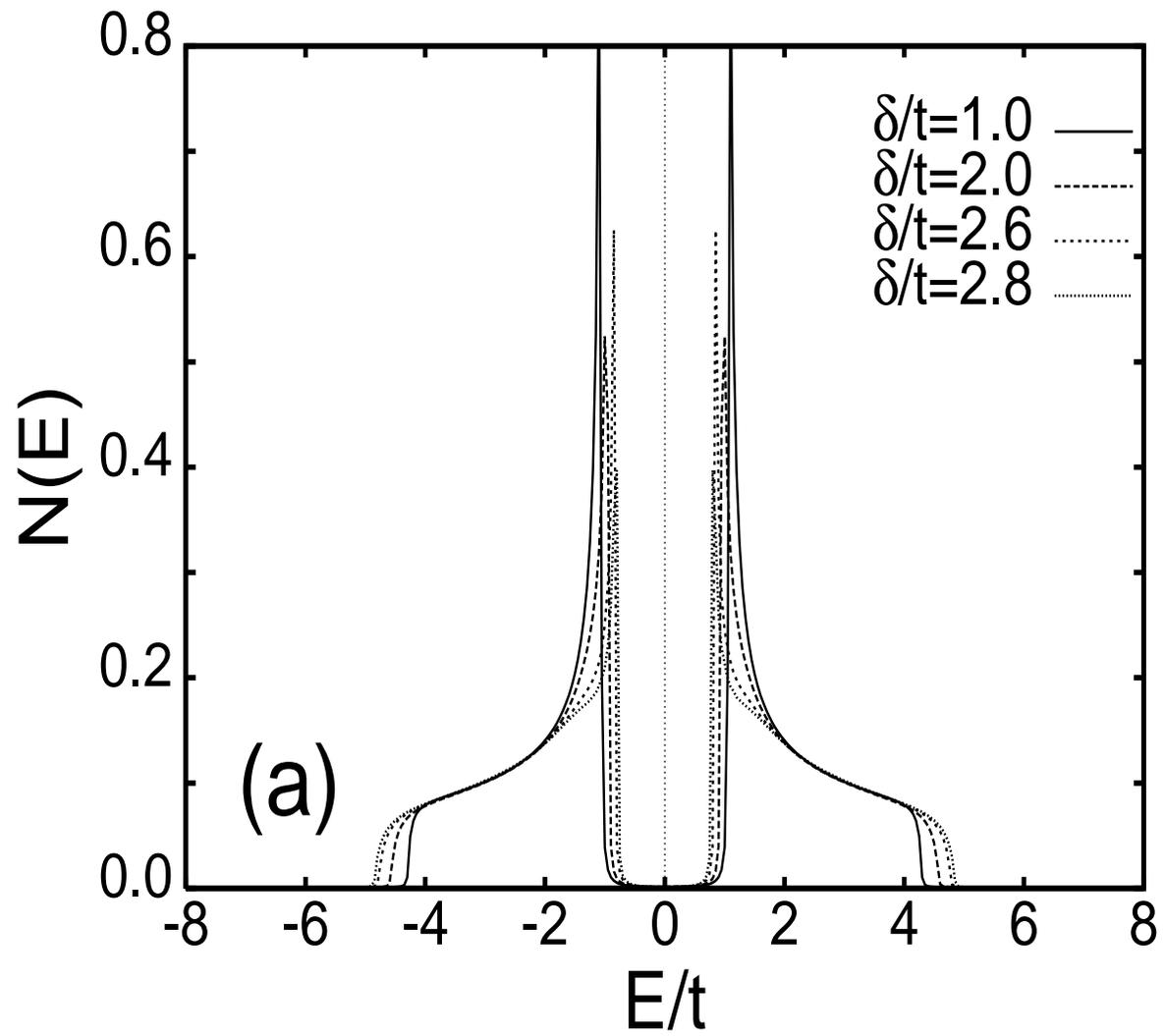
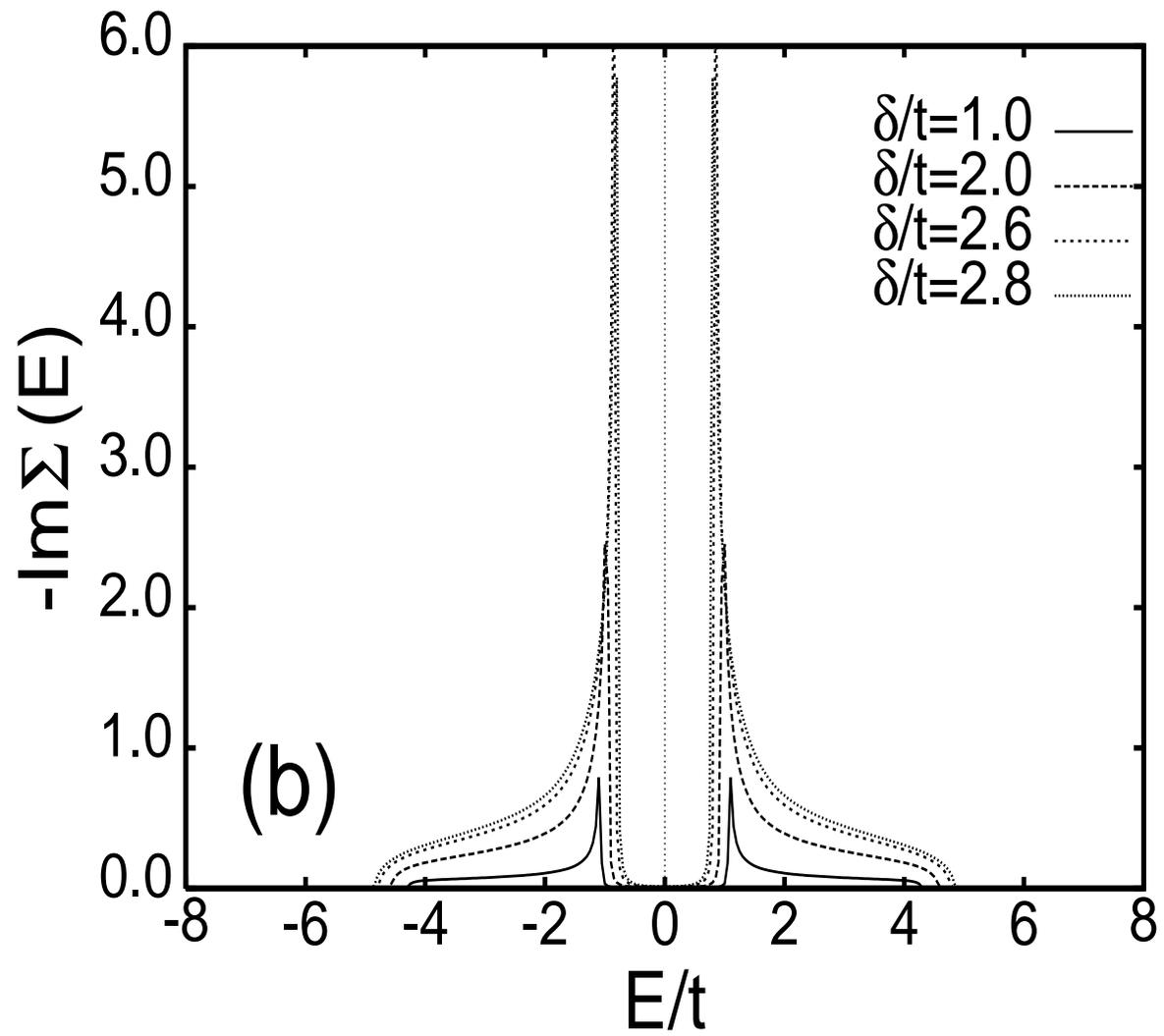
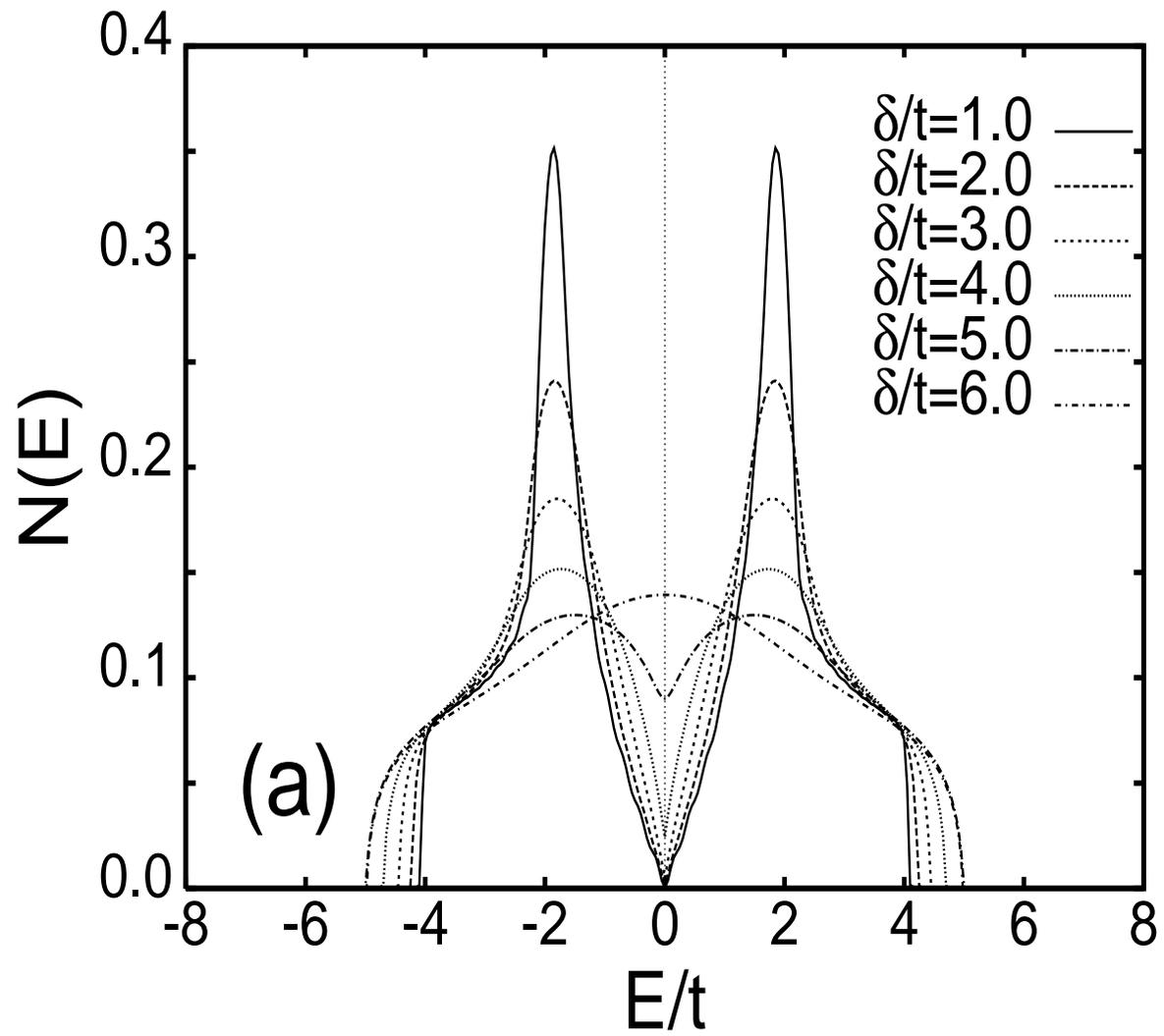
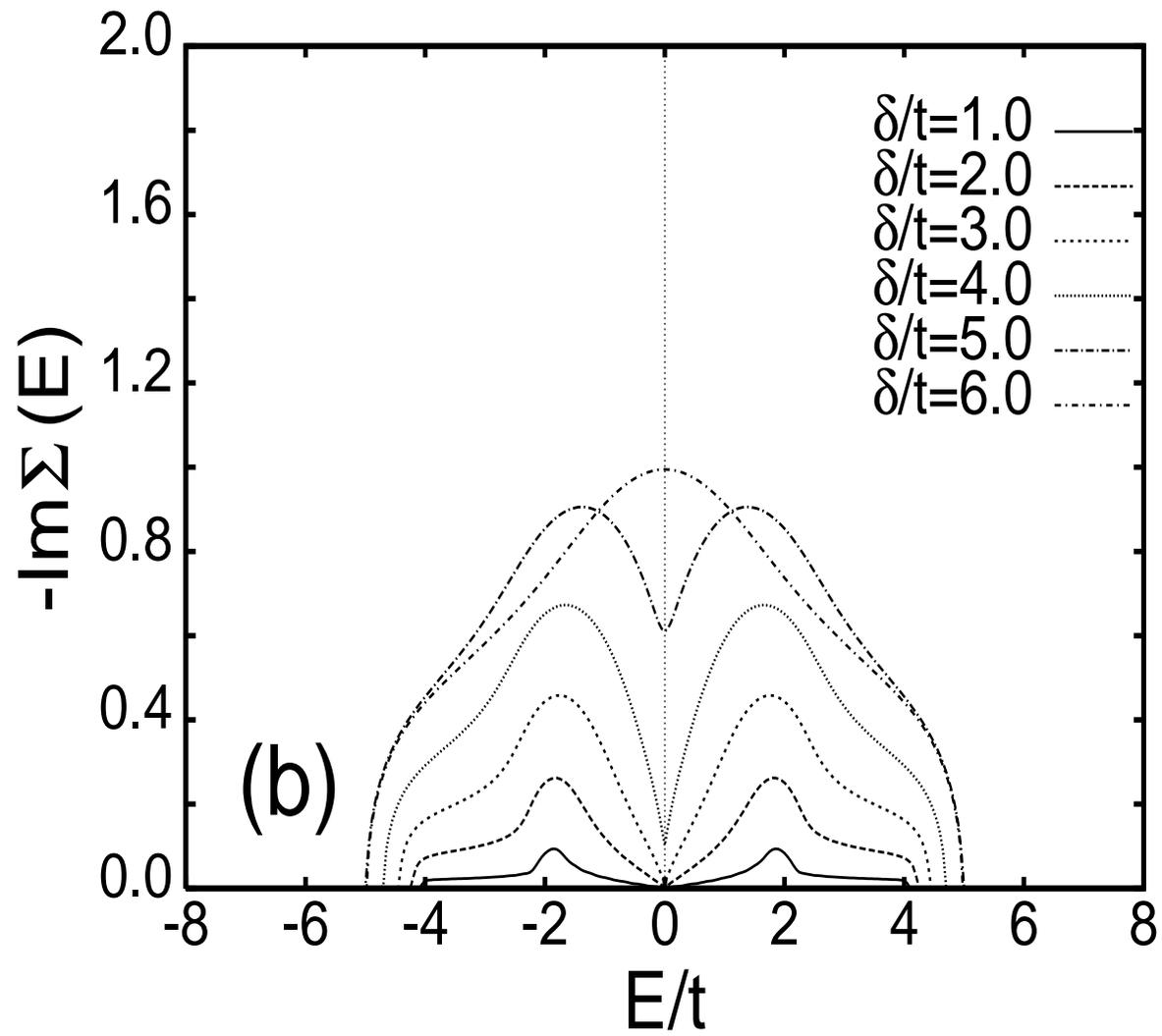
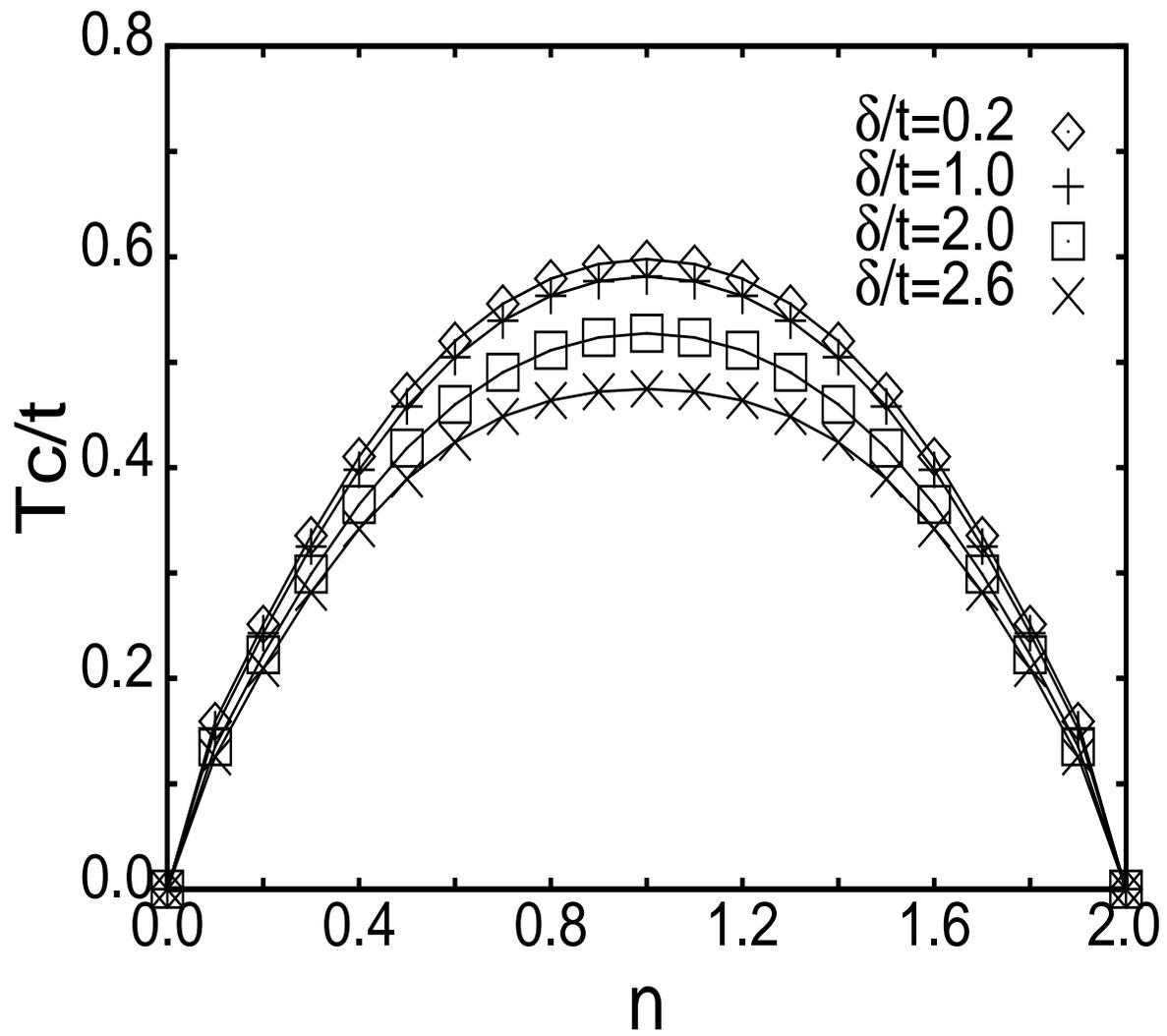
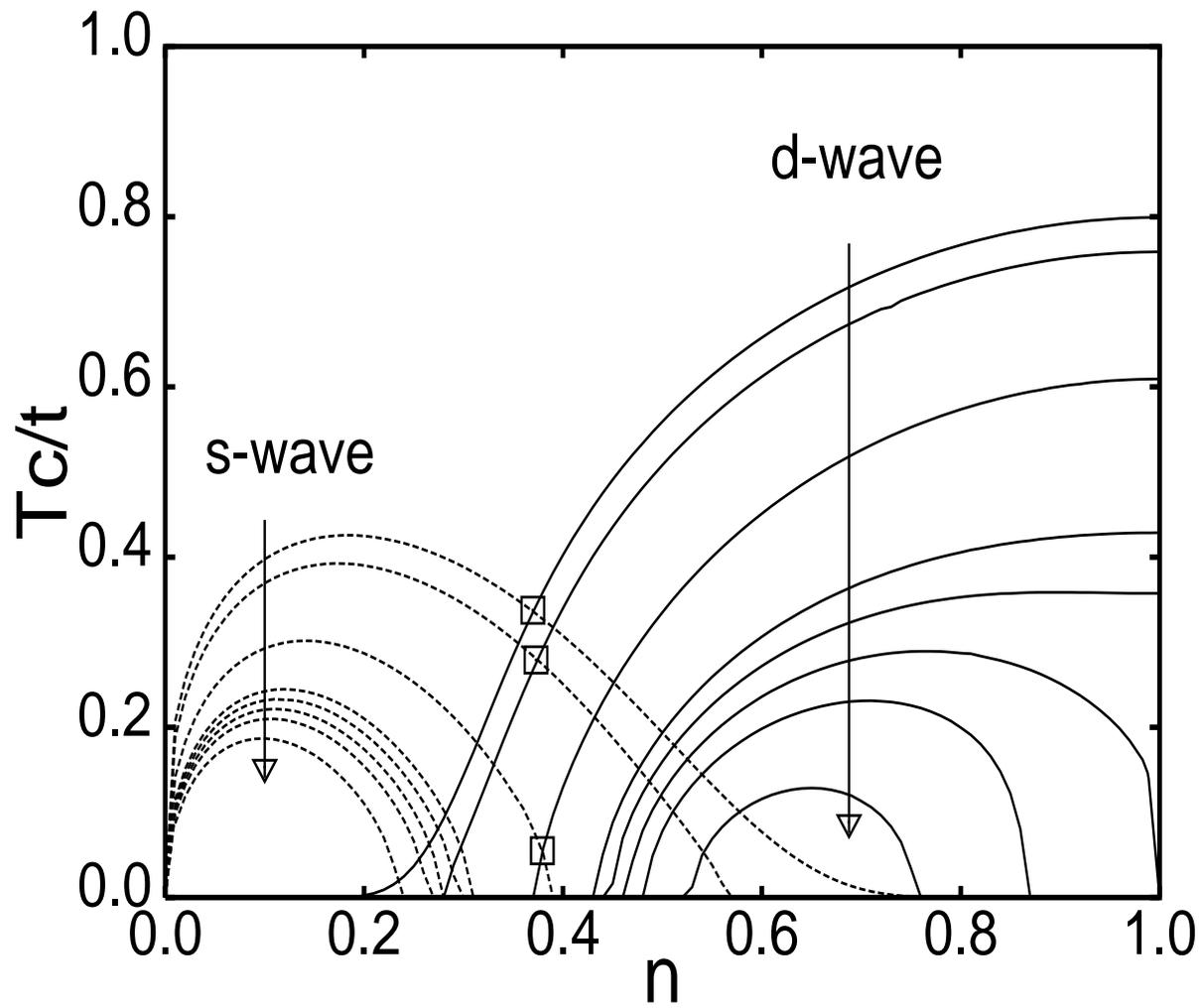
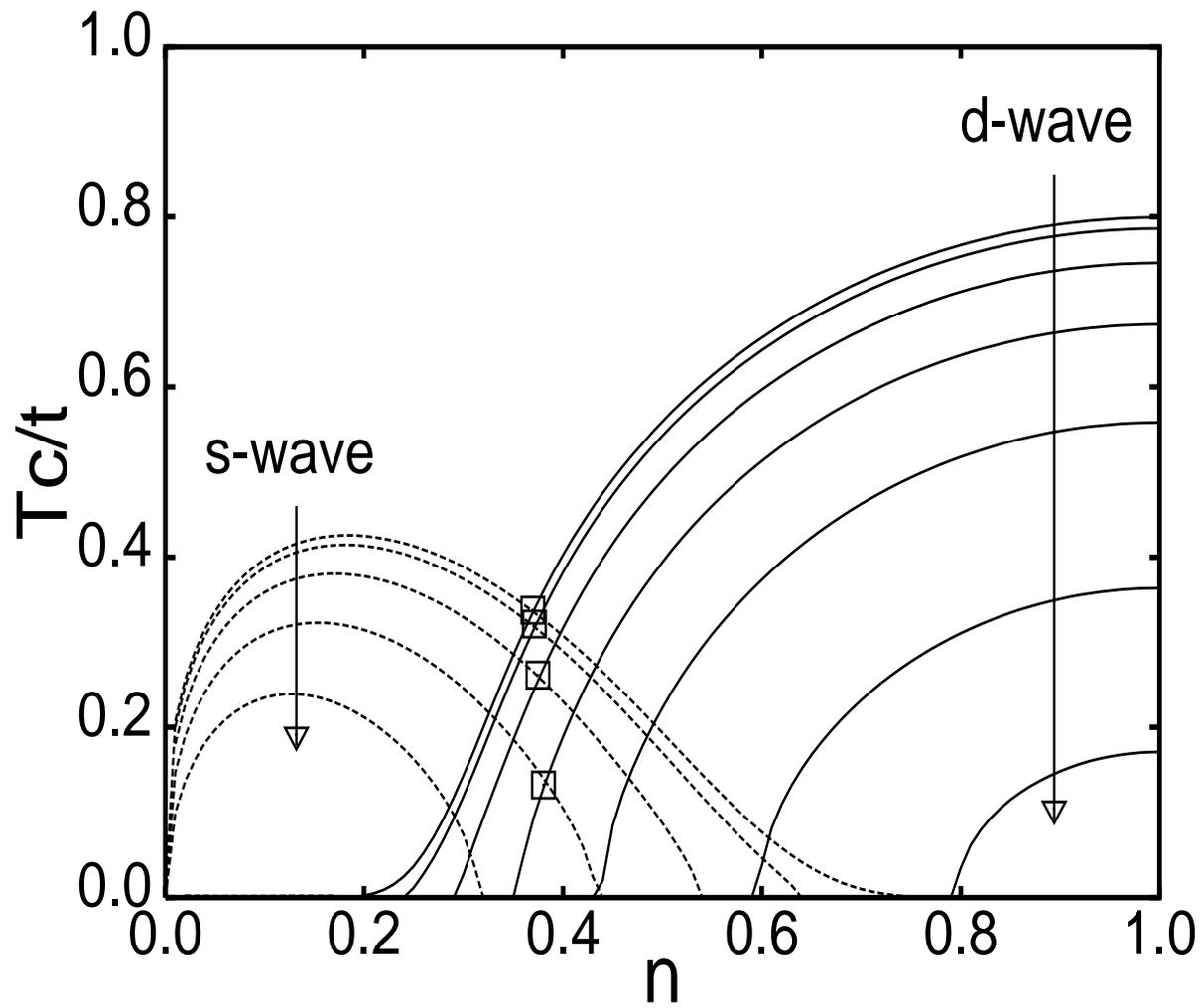
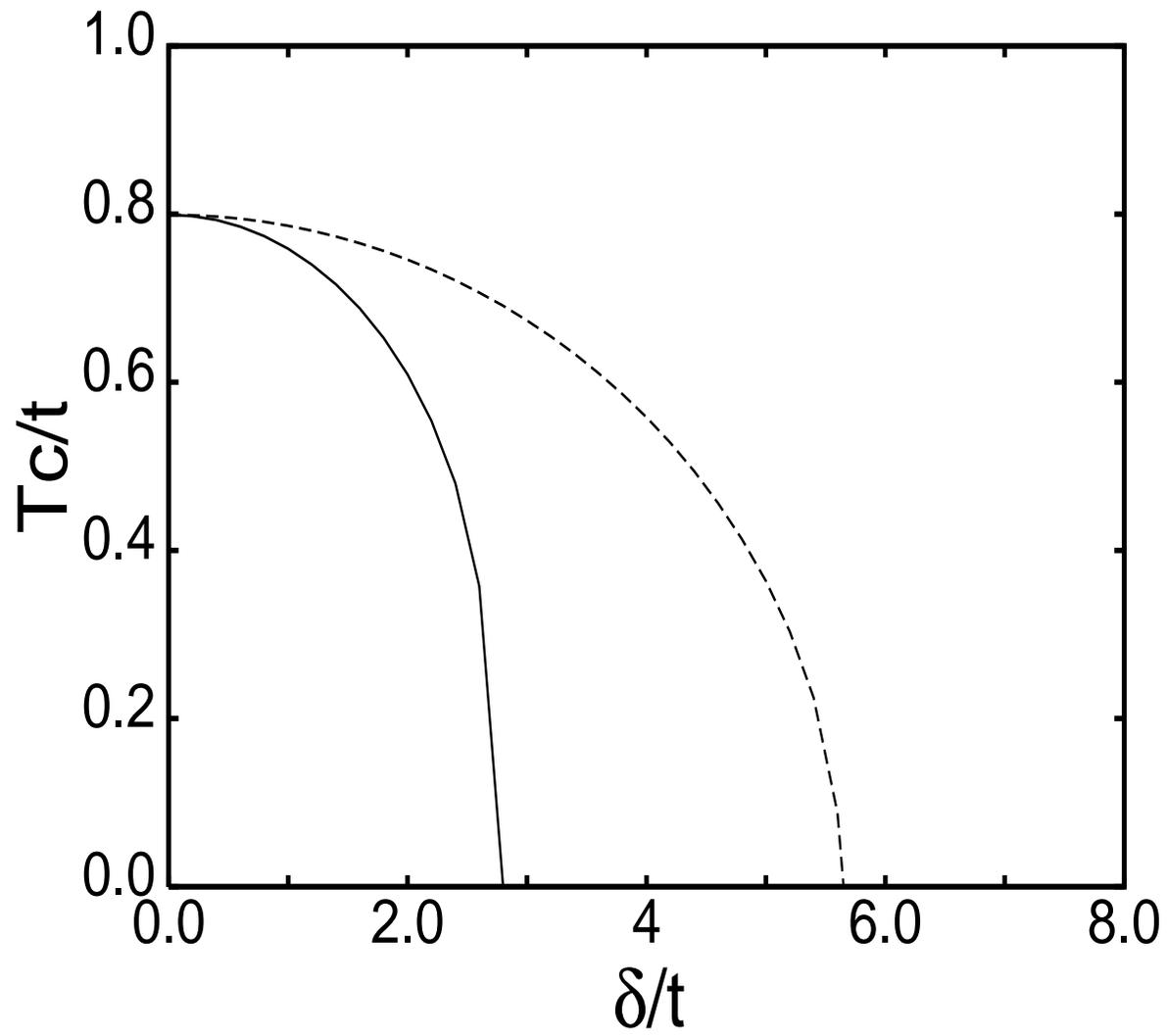
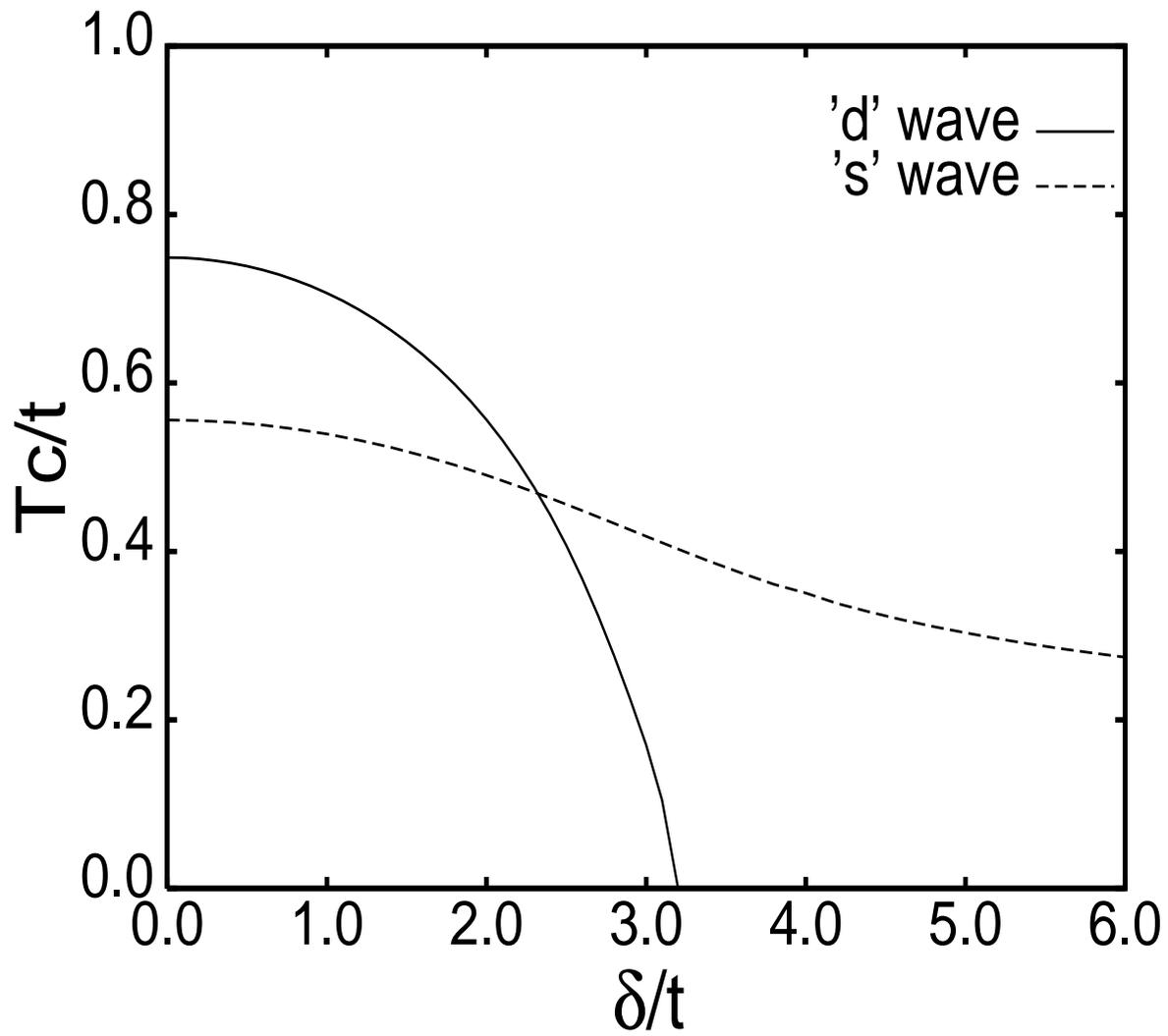
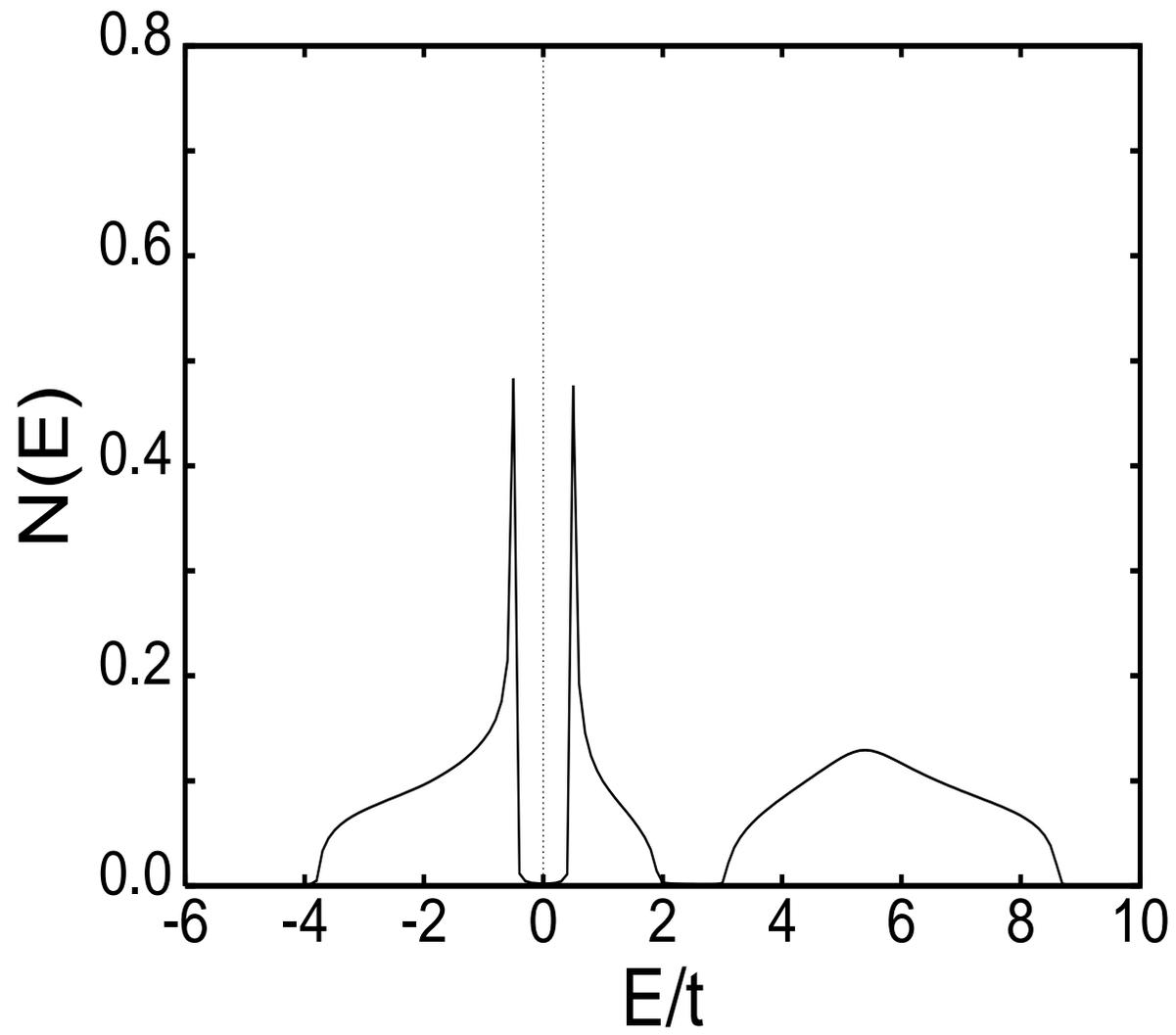
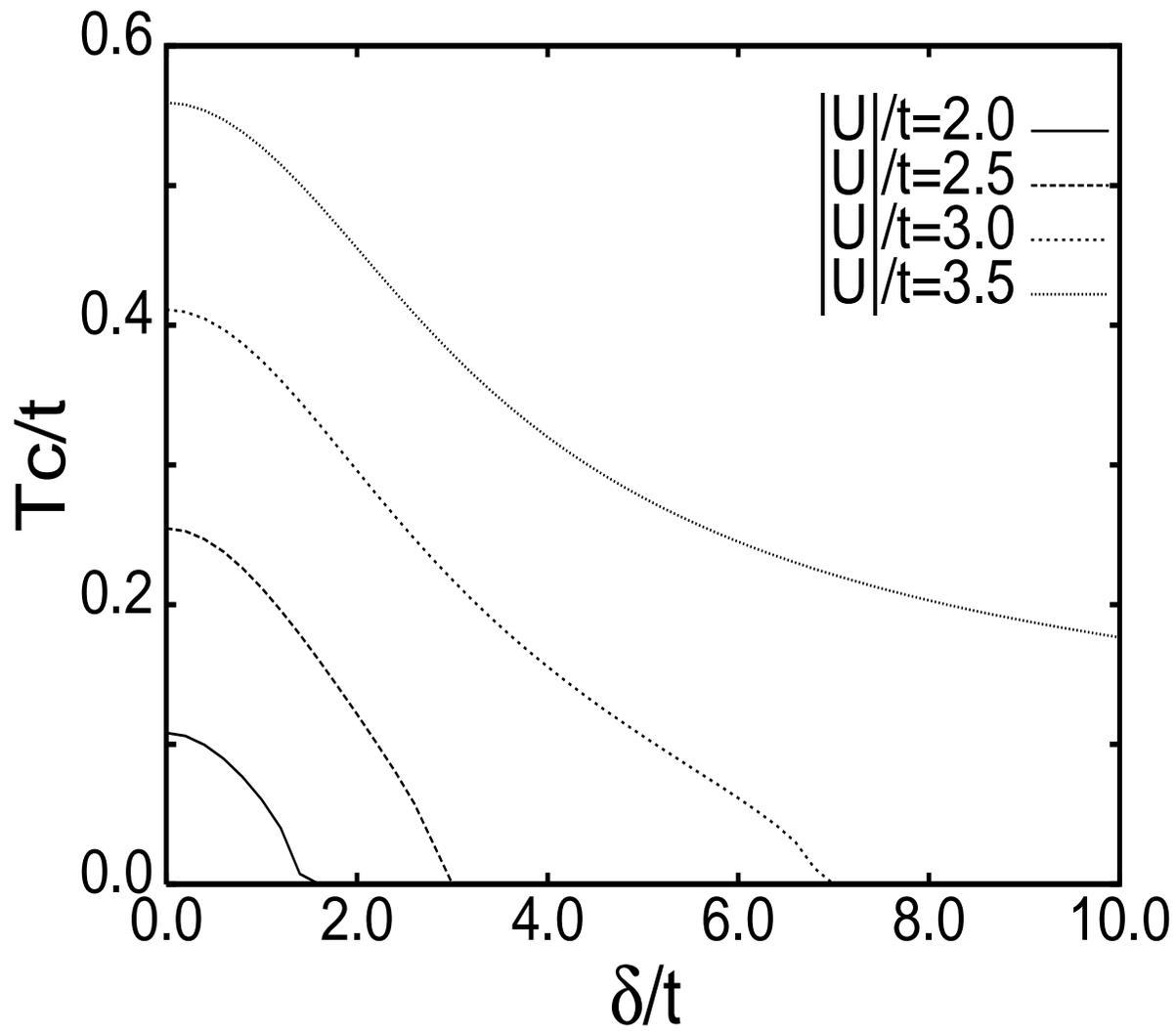
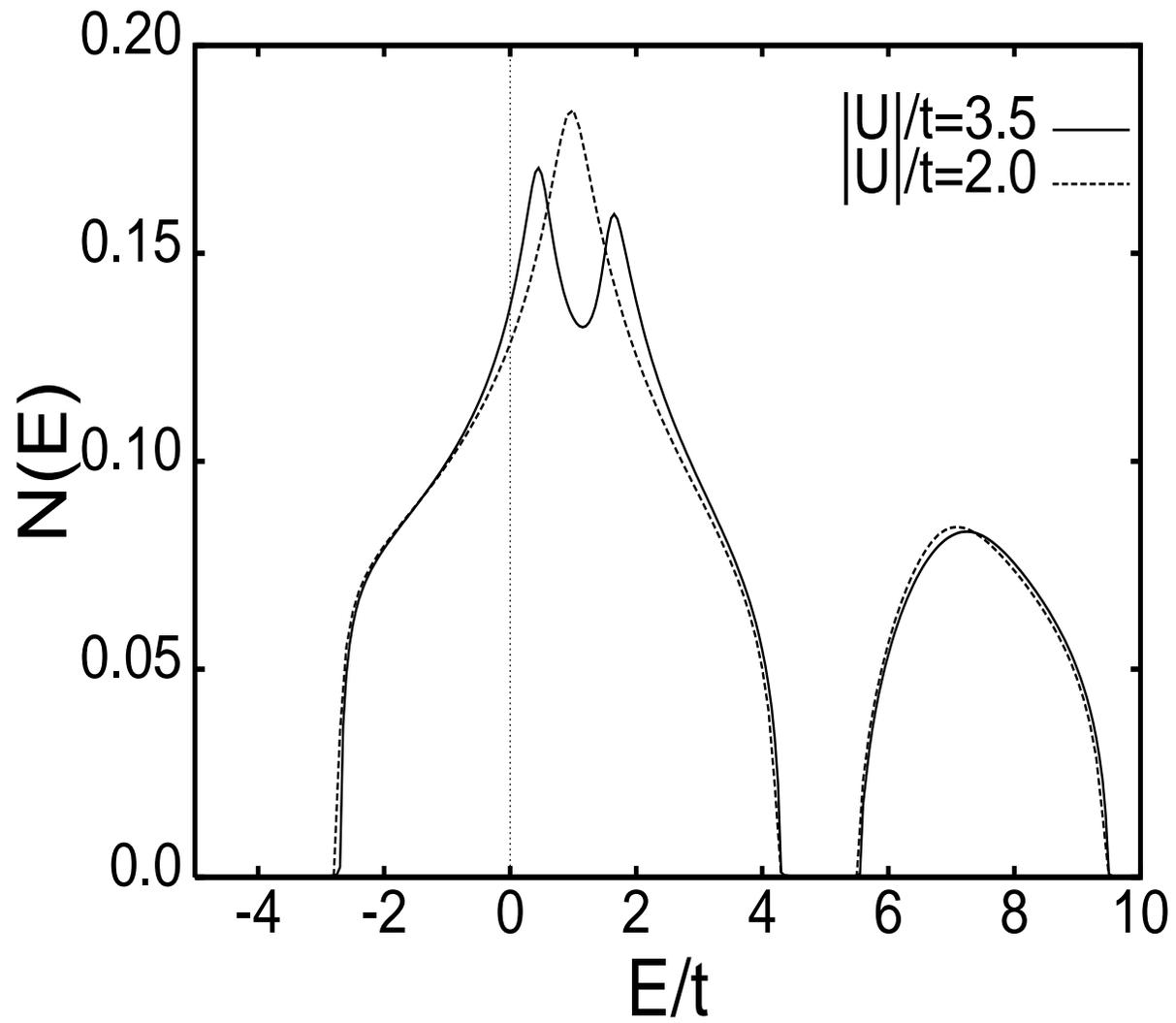
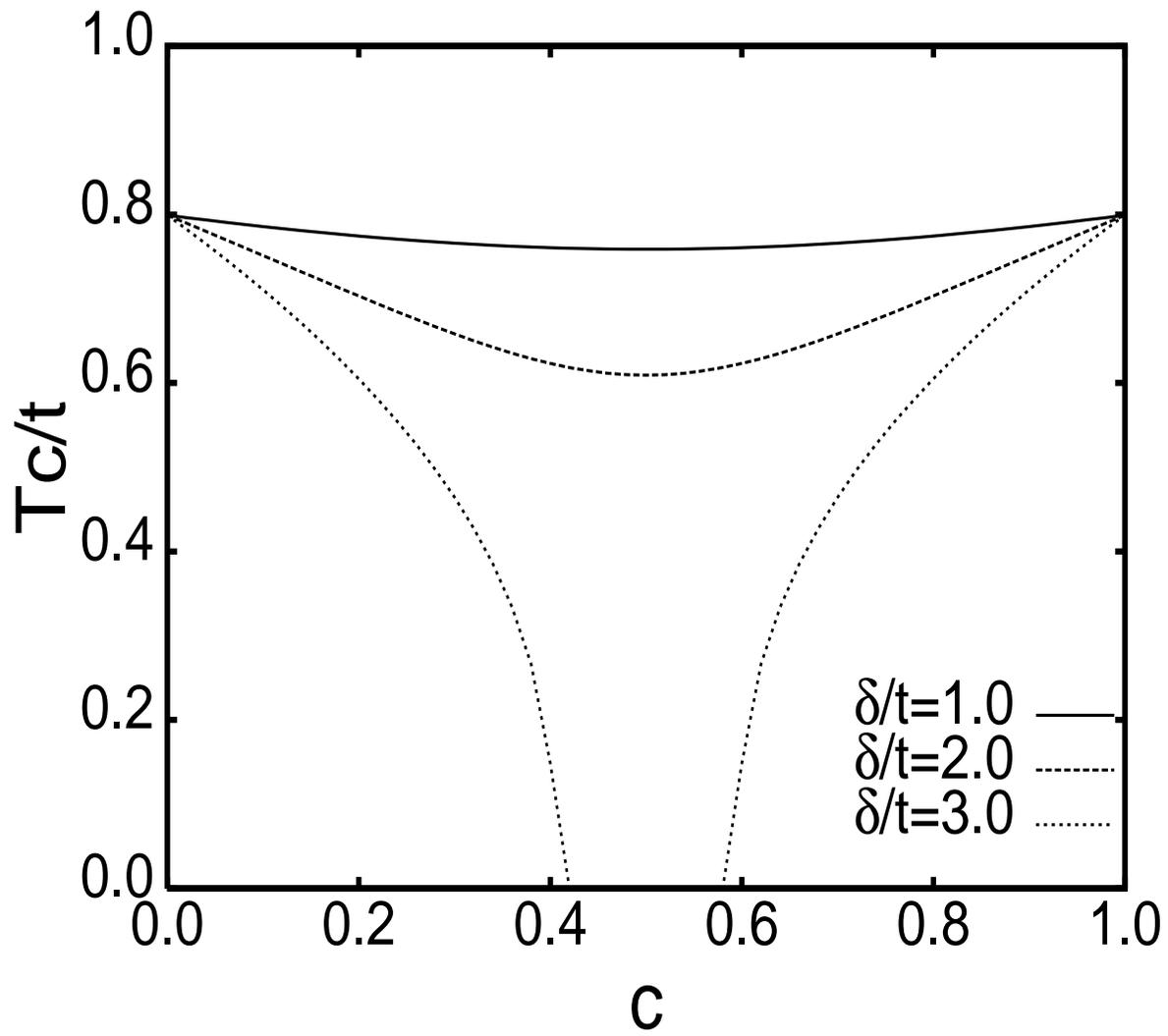
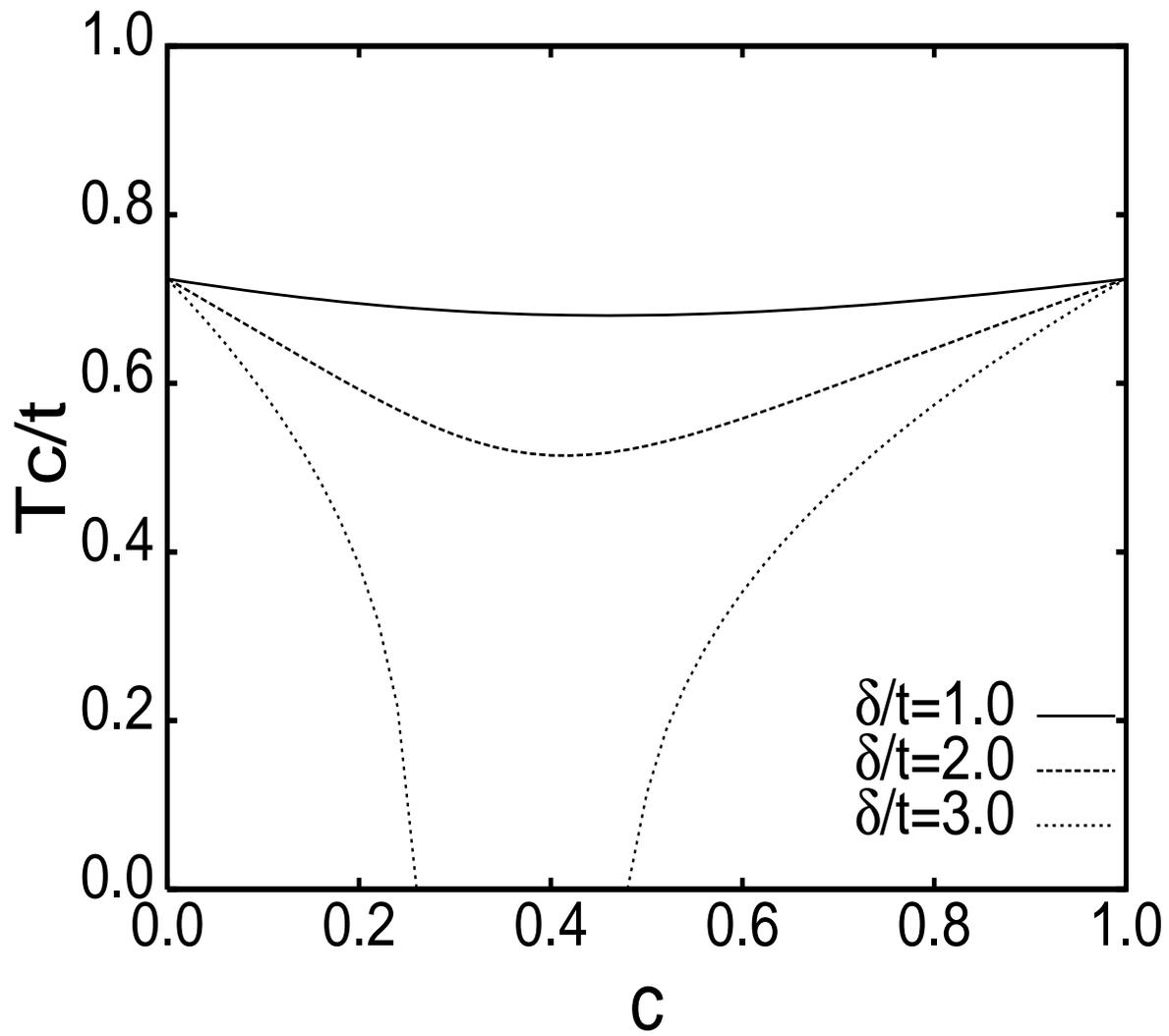
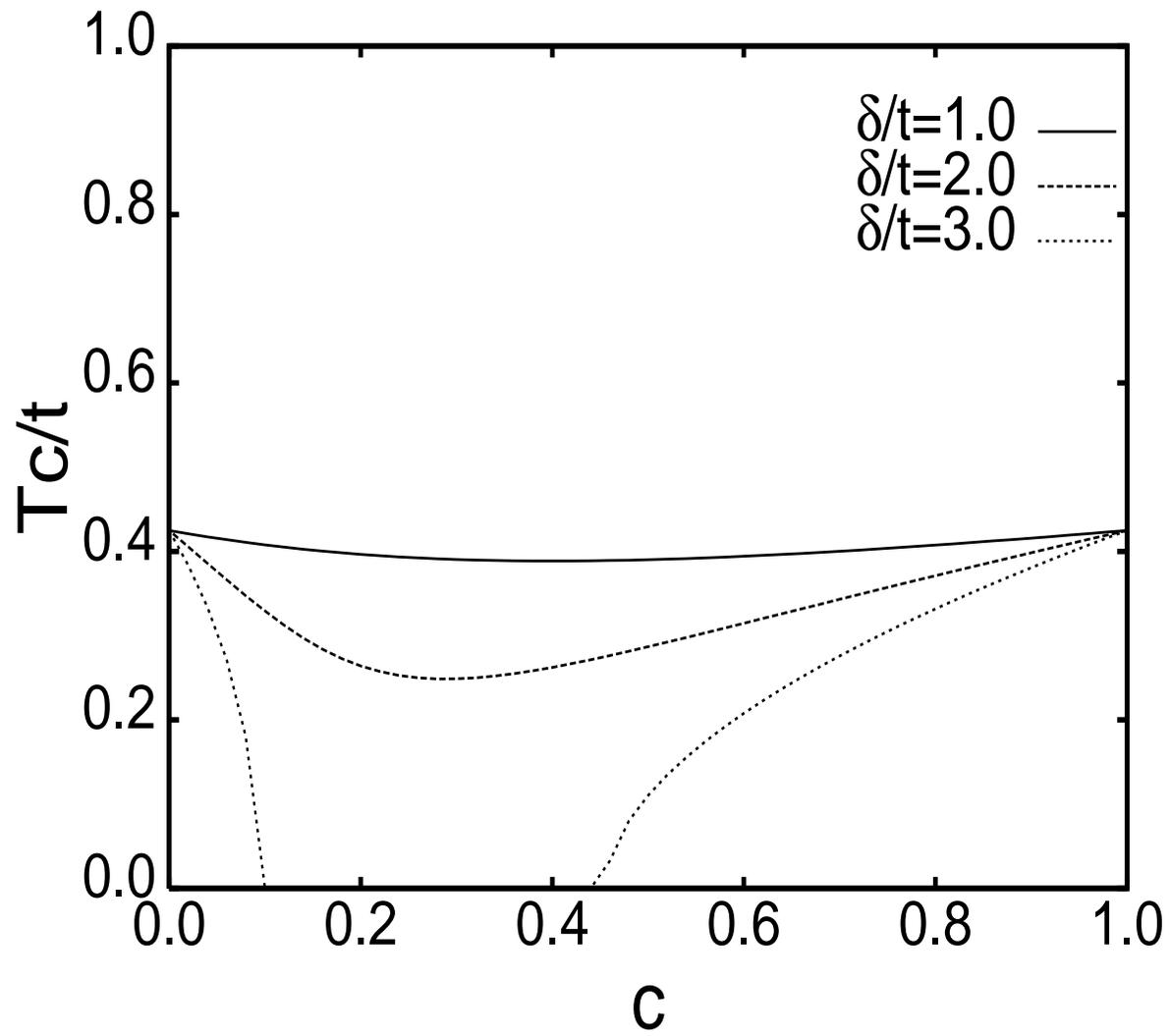
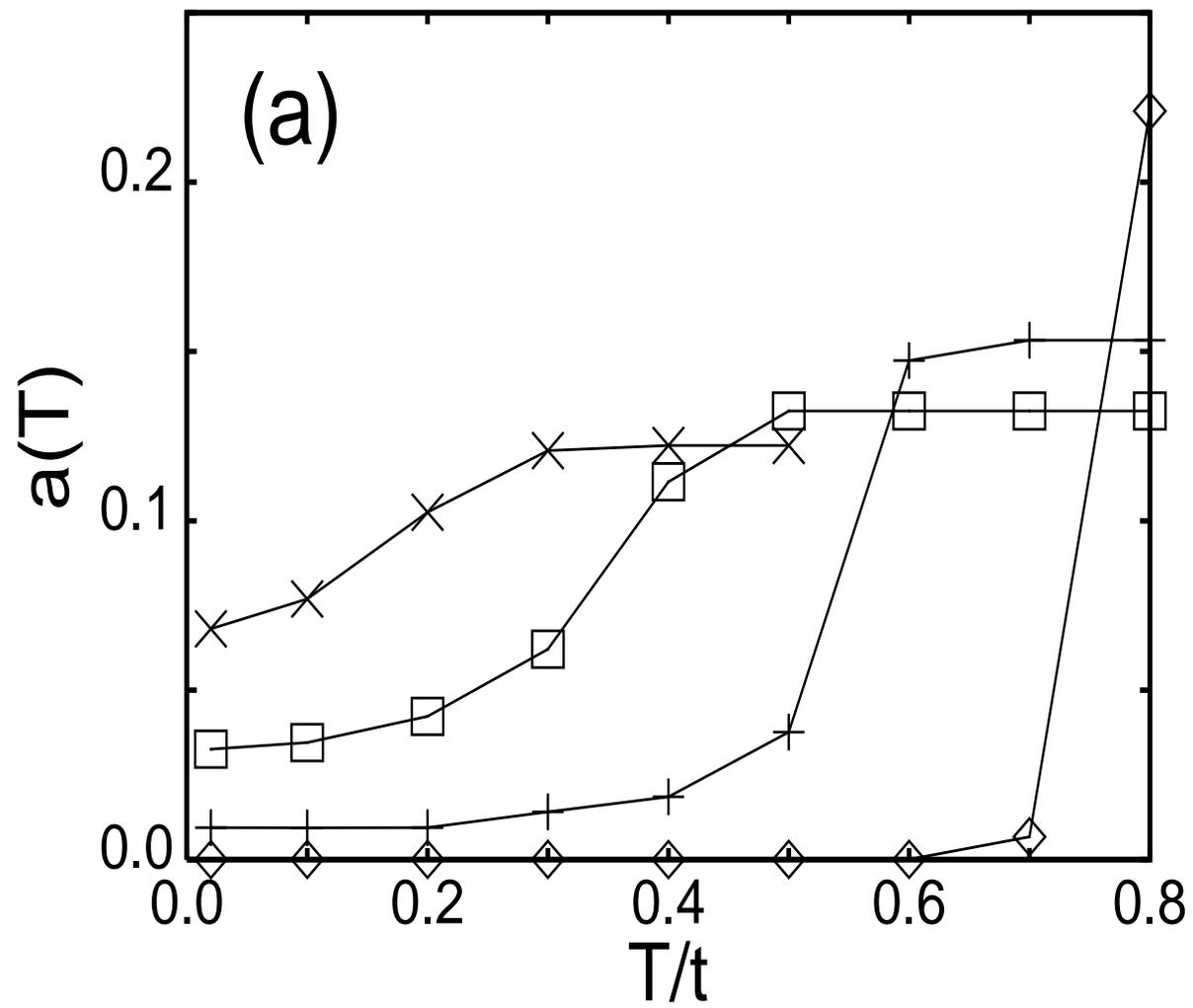
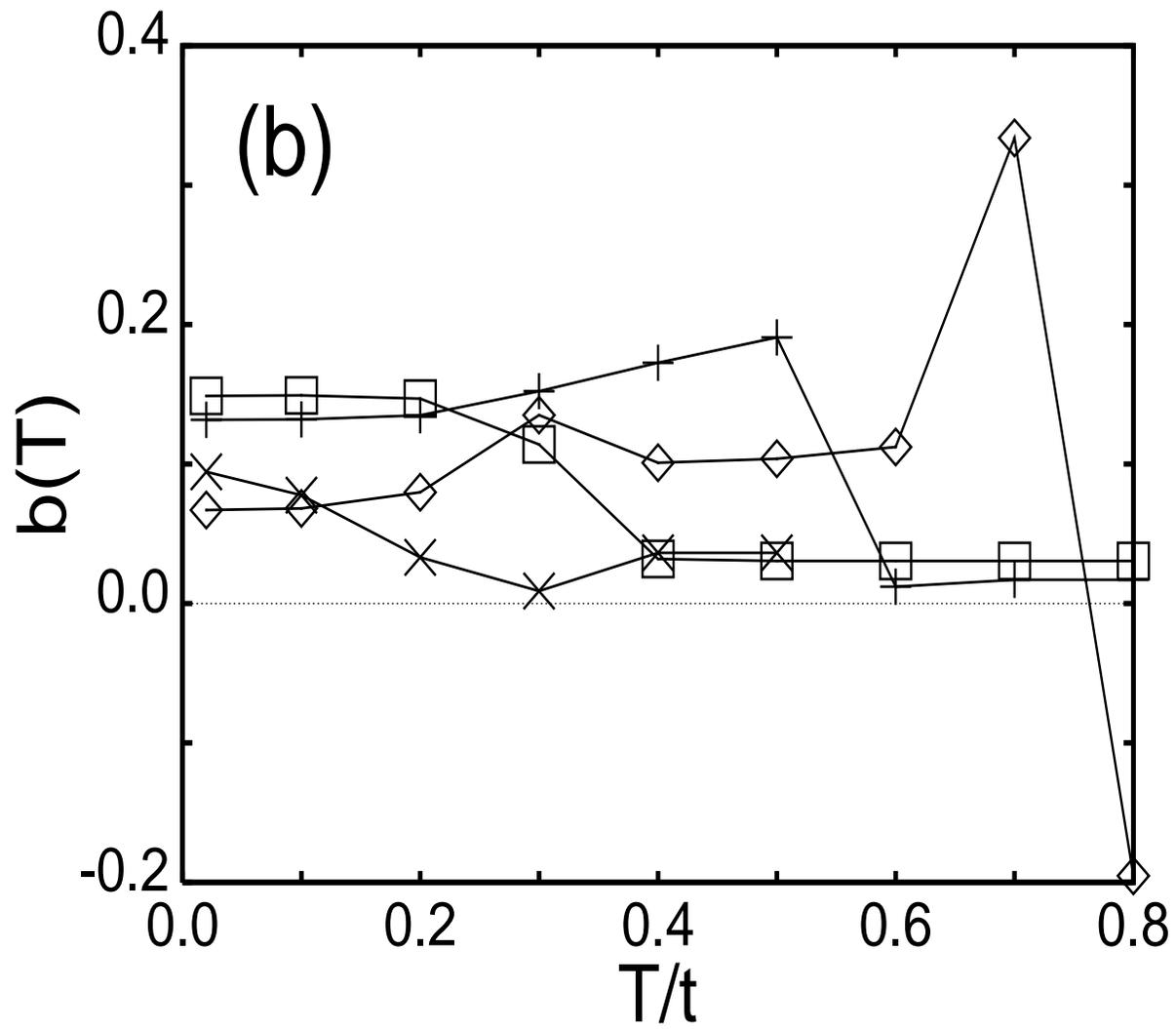
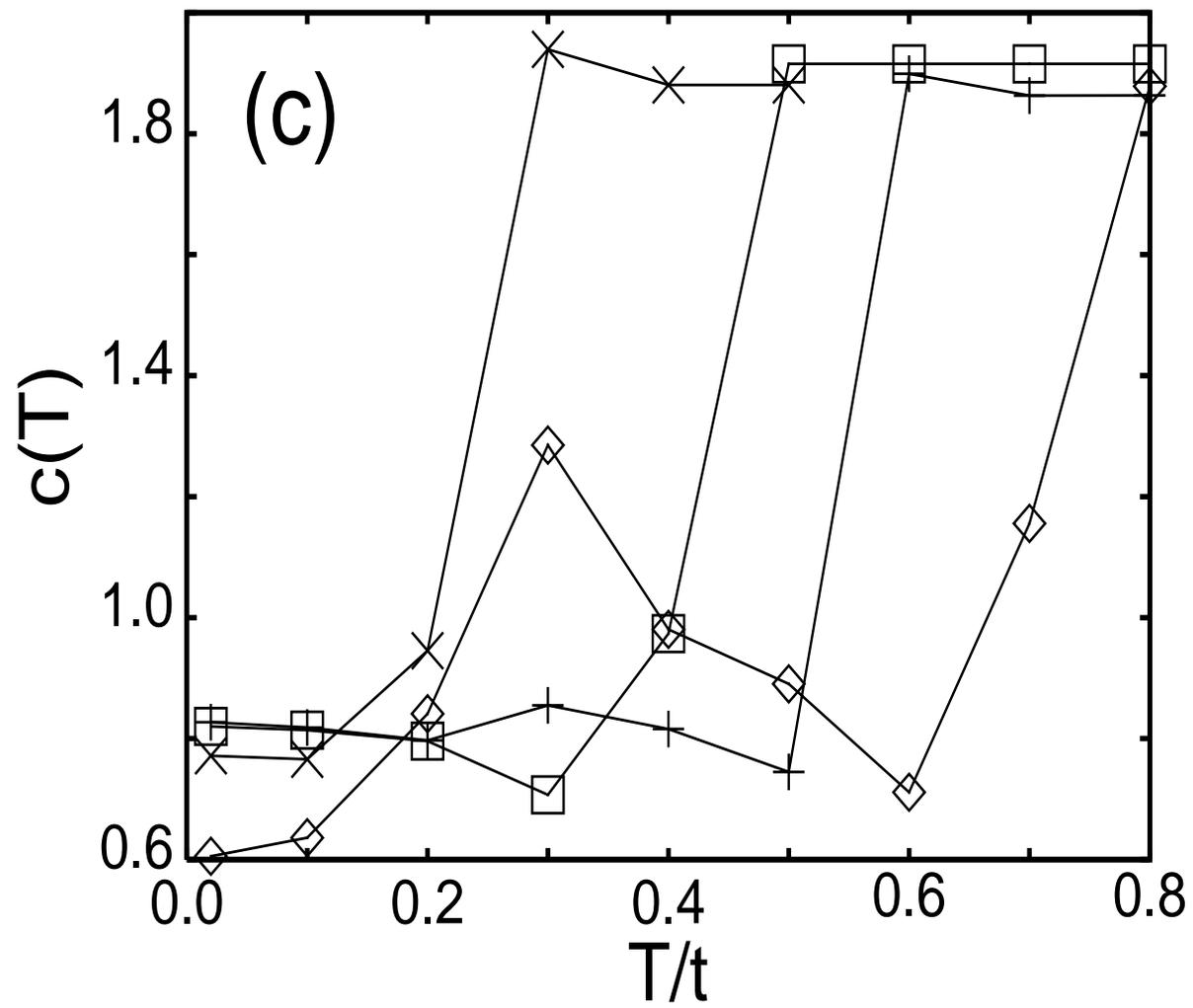
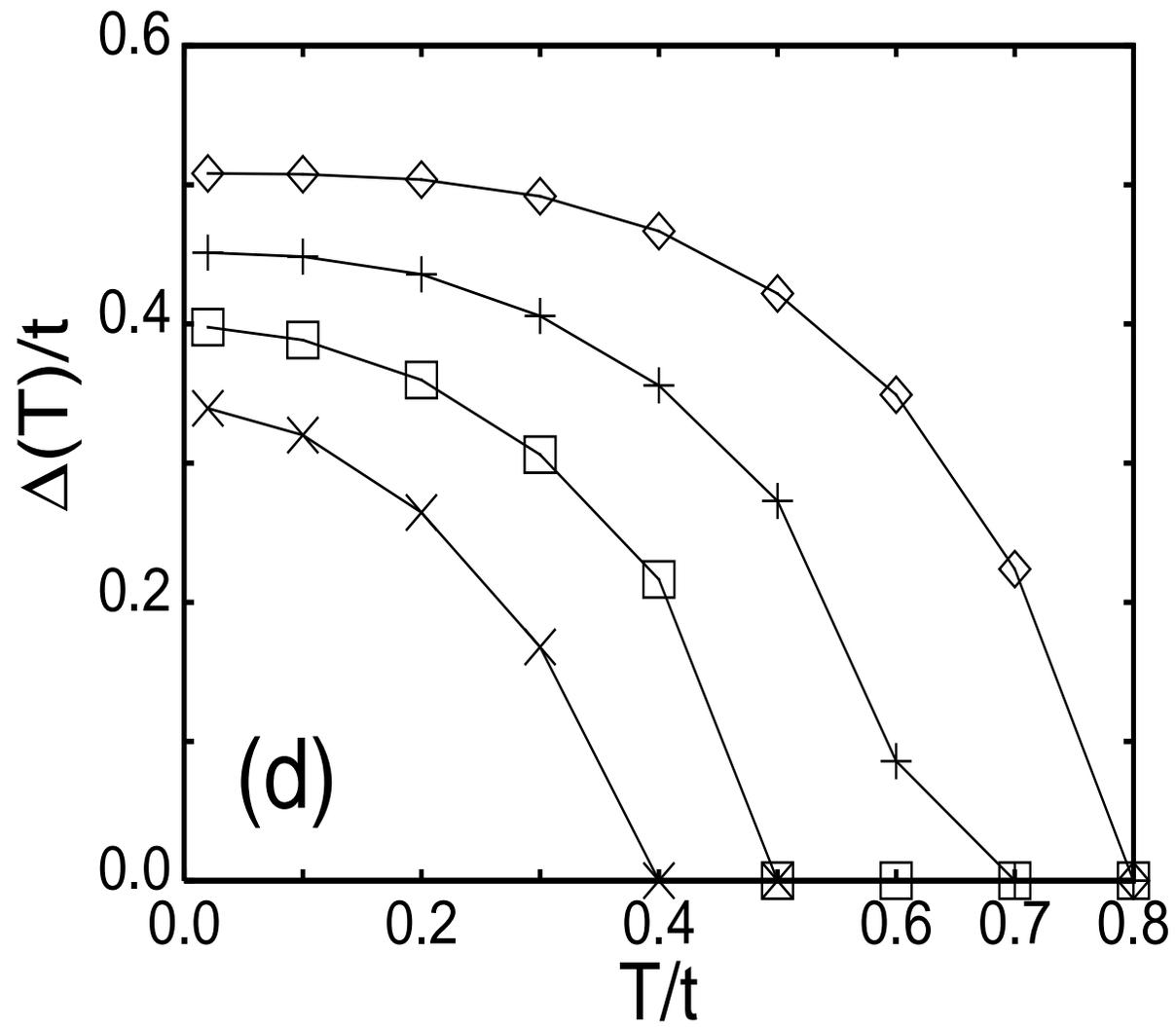
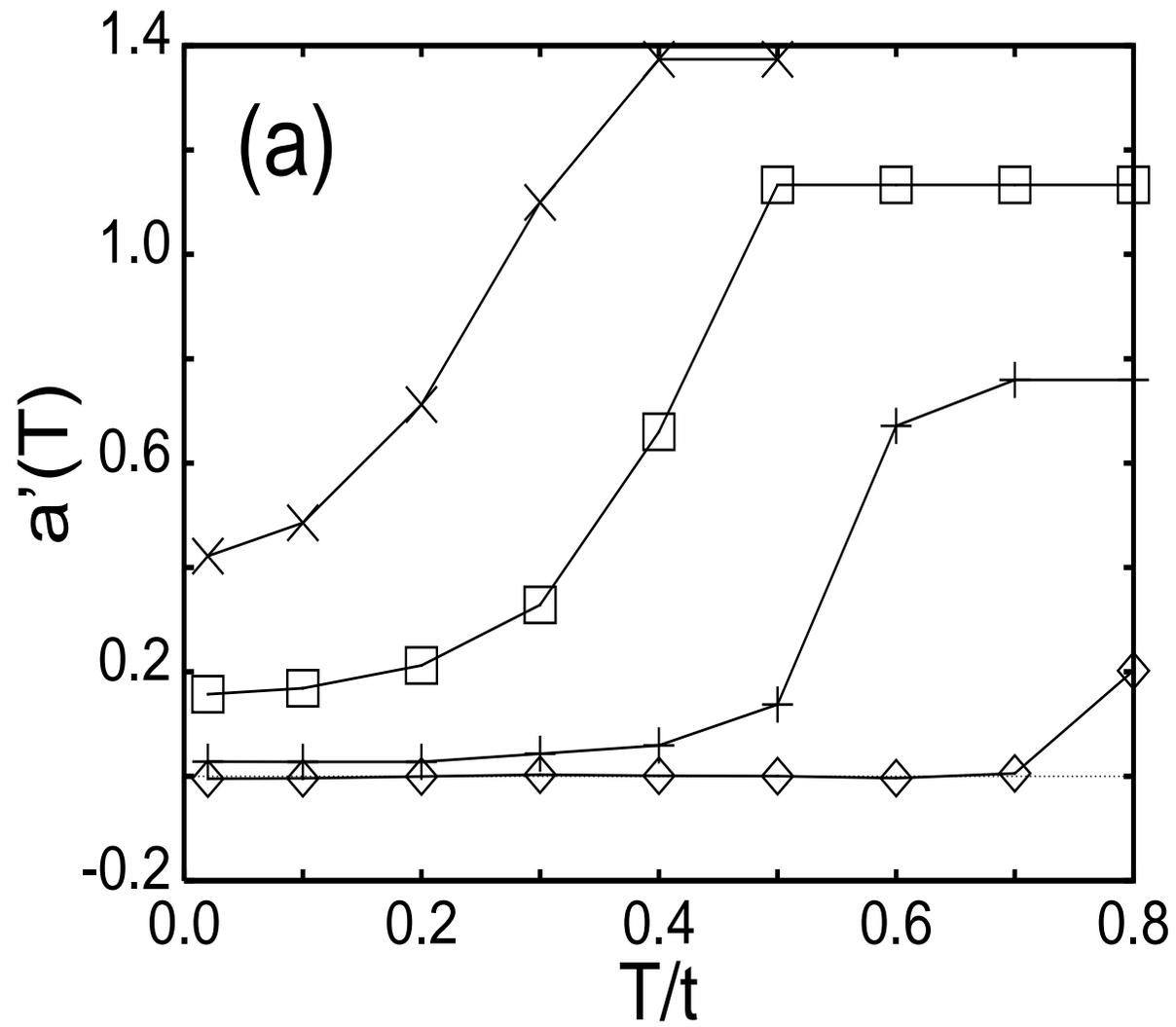
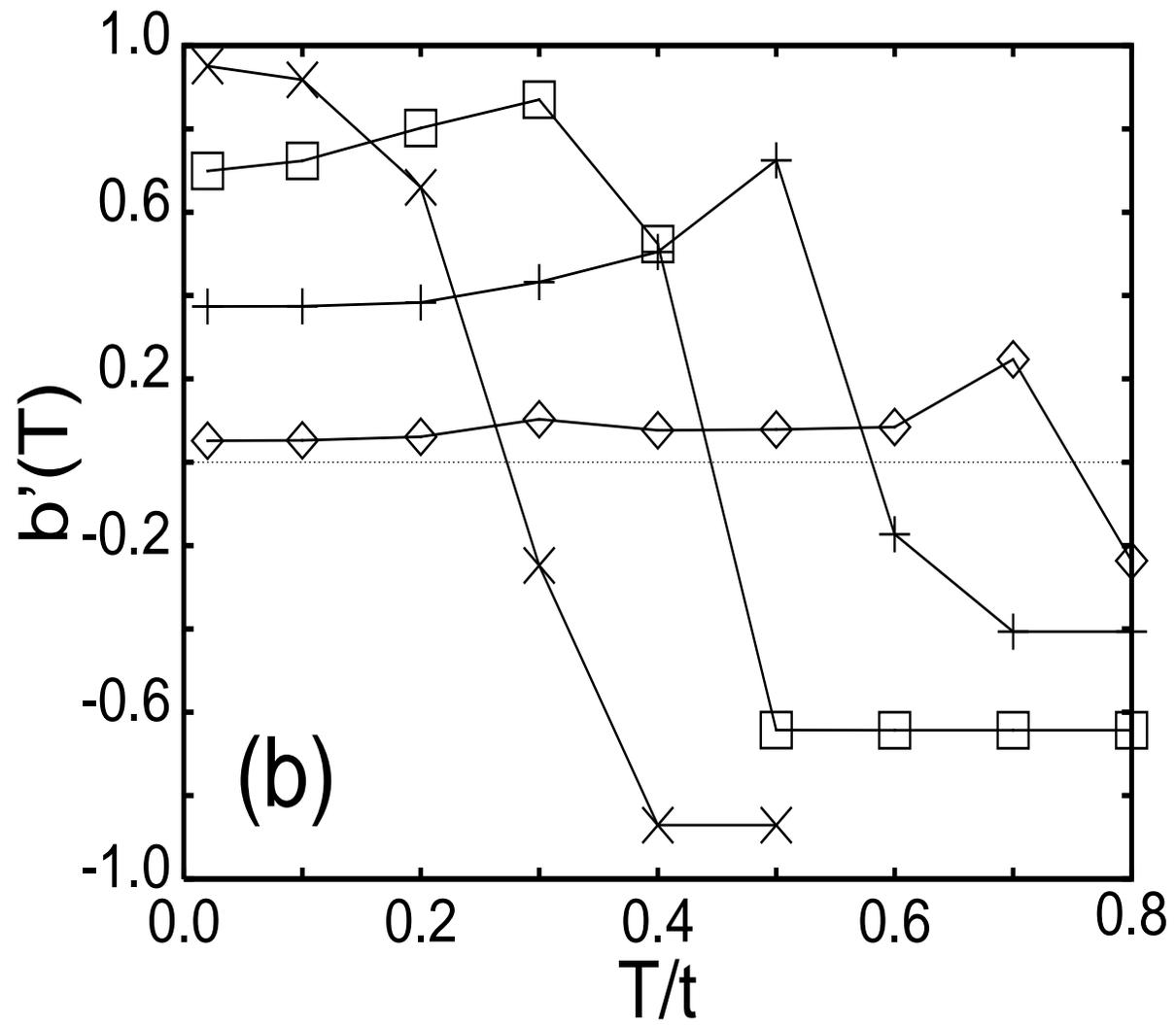
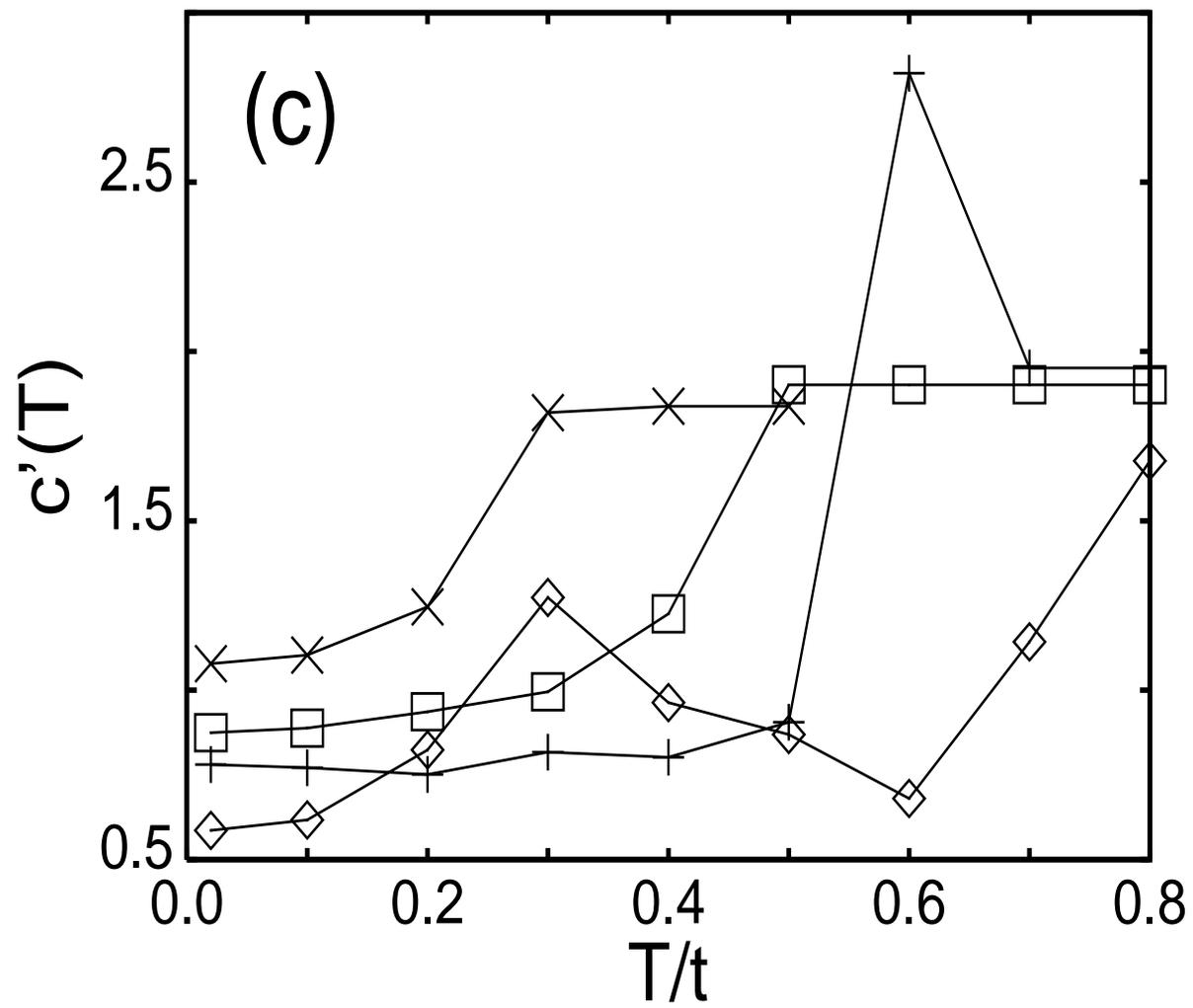
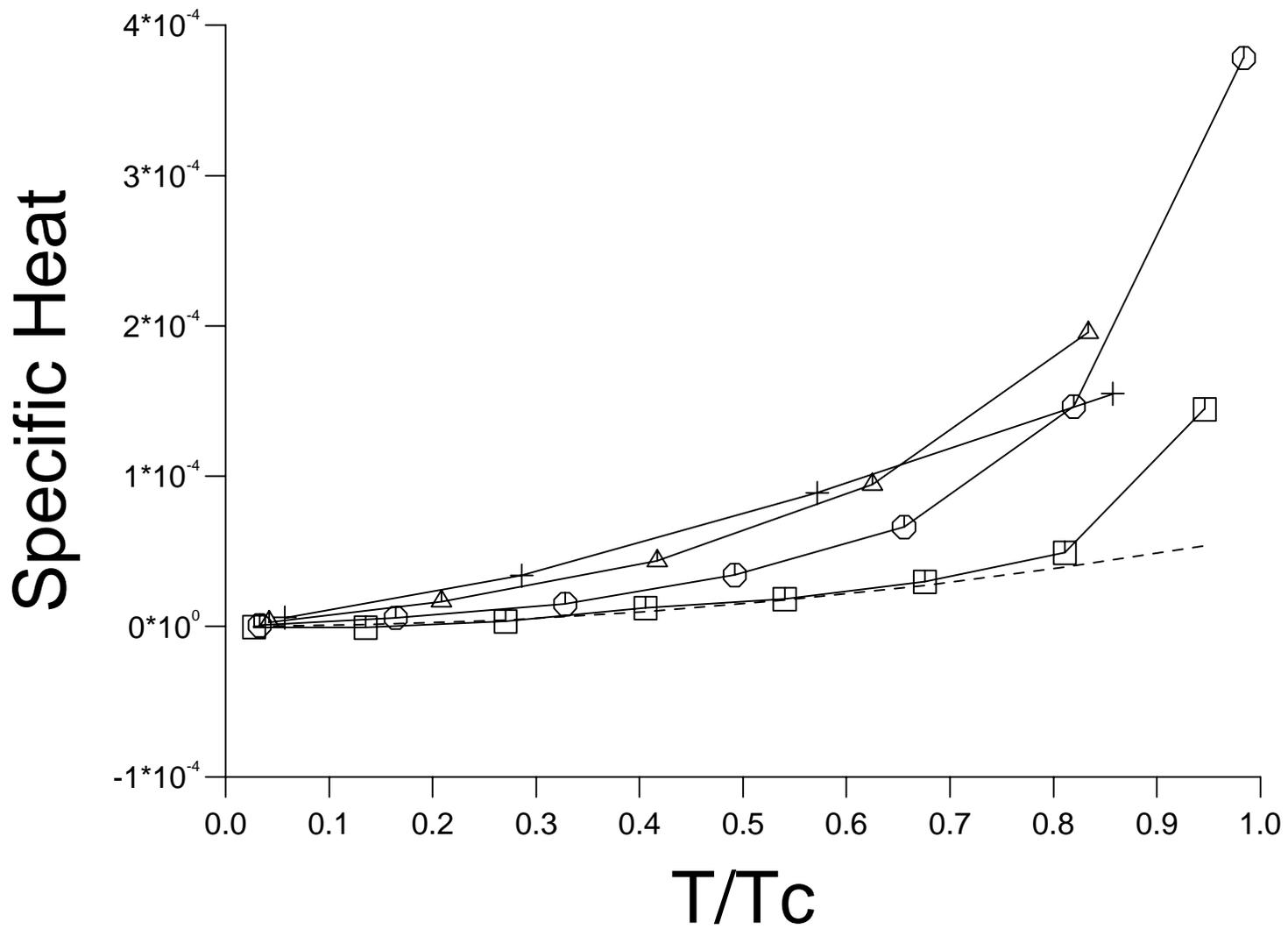
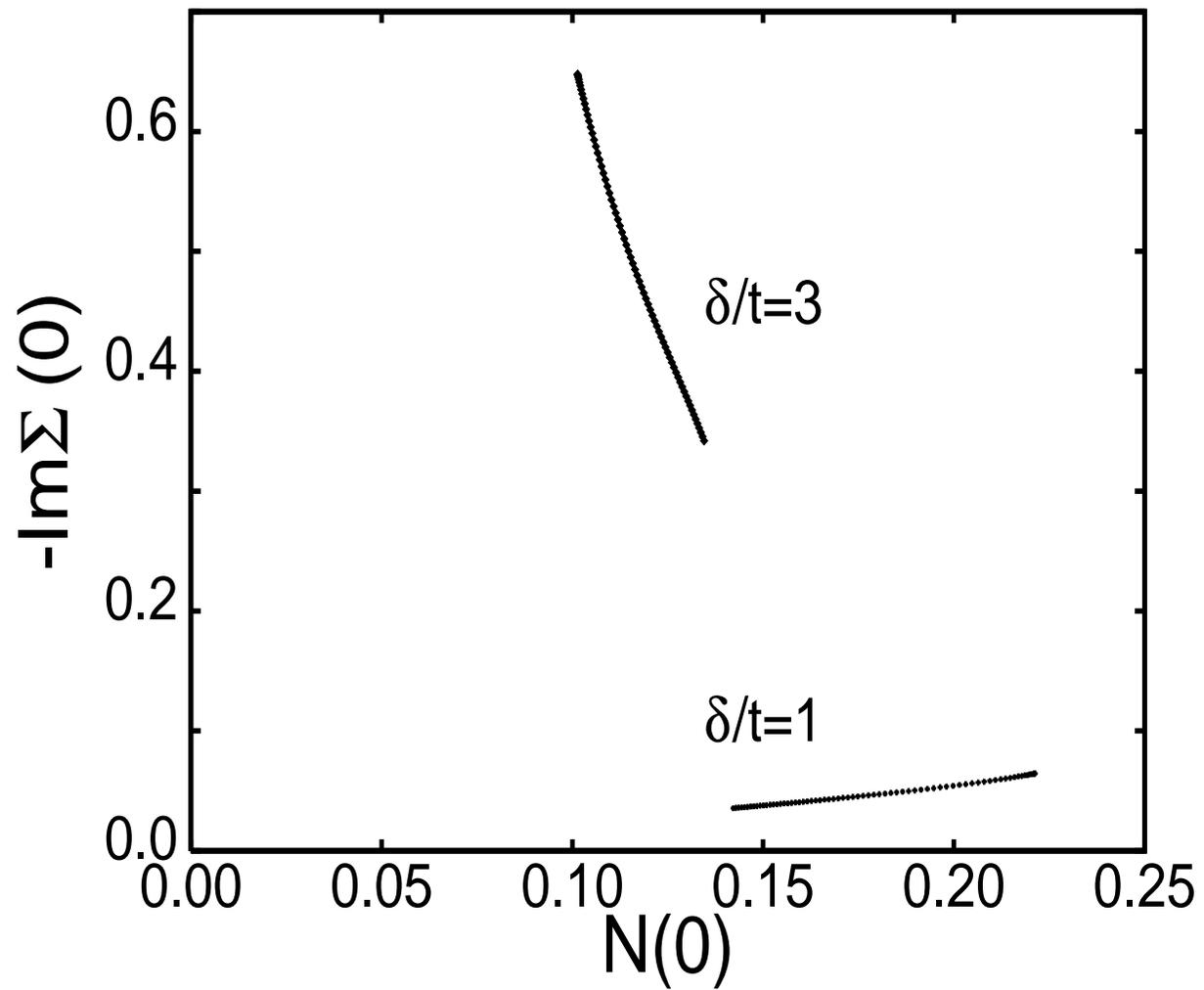
\end{document}